\documentclass[twocolumn,showpacs,preprintnumbers,amsmath,amssymb,superscriptaddress,amssymb]{revtex4-2}
\usepackage{graphicx}% Include figure files
\usepackage{epsfig}
\usepackage{multirow}
\usepackage{pbox}
\usepackage{array}
\usepackage{relsize}
\usepackage{siunitx}
\usepackage{braket}
\usepackage{color,soul}
\usepackage{longtable}

\begin{document}
\newcolumntype{P}[1]{>{\centering\arraybackslash}p{#1}}

\title{One-neutron transfer reaction in the $^{18}$O + $^{48}$Ti collision at 275 MeV}% Force line breaks with \\

\author{O. Sgouros}
\email{onoufrios.sgouros@lns.infn.it}
\affiliation{Dipartimento di Fisica e Astronomia "Ettore Majorana", Universit\`a di Catania, Catania, Italy}
\affiliation{INFN - Laboratori Nazionali del Sud, Catania, Italy}
\author{M. Cutuli}
\affiliation{Dipartimento di Fisica e Astronomia "Ettore Majorana", Universit\`a di Catania, Catania, Italy}
\affiliation{INFN - Laboratori Nazionali del Sud, Catania, Italy}
\author{F. Cappuzzello}
\affiliation{Dipartimento di Fisica e Astronomia "Ettore Majorana", Universit\`a di Catania, Catania, Italy}
\affiliation{INFN - Laboratori Nazionali del Sud, Catania, Italy}
\author{M. Cavallaro}
\affiliation{INFN - Laboratori Nazionali del Sud, Catania, Italy}
\author{D. Carbone}
\affiliation{INFN - Laboratori Nazionali del Sud, Catania, Italy}
\author{C. Agodi}
\affiliation{INFN - Laboratori Nazionali del Sud, Catania, Italy}
\author{G. De Gregorio}
\affiliation{INFN - Sezione di Napoli, Napoli, Italy}
\affiliation{Dipartimento di Matematica e Fisica, Universit\`a della Campania "Luigi Vanvitelli", Caserta, Italy}
\author{A. Gargano}
\affiliation{INFN - Sezione di Napoli, Napoli, Italy}
\author{R. Linares}
\affiliation{Instituto de F\'isica, Universidade Federal Fluminense, Niter\'oi, Brazil}
\author{G. A. Brischetto}
\affiliation{Dipartimento di Fisica e Astronomia "Ettore Majorana", Universit\`a di Catania, Catania, Italy}
\affiliation{INFN - Laboratori Nazionali del Sud, Catania, Italy}
\author{D. Calvo}
\affiliation{INFN - Sezione di Torino, Torino, Italy}
\author{E. R. Ch\'avez Lomel\'i}
\affiliation{Instituto de F\'isica, Universidad Nacional Aut\'onoma de M\'exico, Mexico City, Mexico}
\author{I. Ciraldo}
\affiliation{Dipartimento di Fisica e Astronomia "Ettore Majorana", Universit\`a di Catania, Catania, Italy}
\affiliation{INFN - Laboratori Nazionali del Sud, Catania, Italy}
\author{F. Delaunay}
\affiliation{Dipartimento di Fisica e Astronomia "Ettore Majorana", Universit\`a di Catania, Catania, Italy}
\affiliation{INFN - Laboratori Nazionali del Sud, Catania, Italy}
\affiliation{LPC Caen UMR6534, Universit\'e de Caen Normandie, ENSICAEN, CNRS/IN2P3, F-14000 Caen, France}
\author{H. Djapo}
\affiliation{Ancara University, Institute of Accelerator Technologies, Turkey}
\author{C. Eke}
\affiliation{Department of Mathematics and Science Education, Faculty of Education, Akdeniz University, Antalya, Turkey}
\author{P. Finocchiaro}
\affiliation{INFN - Laboratori Nazionali del Sud, Catania, Italy}
\author{M. Fisichella}
\affiliation{INFN - Laboratori Nazionali del Sud, Catania, Italy}
\author{M. A. Guazzelli}
\affiliation{Centro Universitario FEI, S\~ao Bernardo do Campo, Brazil}
\author{A. Hacisalihoglu}
\affiliation{Department of Physics, Recep Tayyip Erdogan University, Rize, Turkey}
\author{J. Lubian}
\affiliation{Instituto de F\'isica, Universidade Federal Fluminense, Niter\'oi, Brazil}
\author{N. H. Medina}
\affiliation{Instituto de F\'isica, Universidade de S\~ao Paulo, S\~ao Paulo, Brazil}
\author{M. Moralles}
\affiliation{Instituto de Pesquisas Energeticas e Nucleares IPEN/CNEN, S\~ao Paulo, Brazil}
\author{J. R. B. Oliveira}
\affiliation{Instituto de F\'isica, Universidade de S\~ao Paulo, S\~ao Paulo, Brazil}
\author{A. Pakou}
\affiliation{Department of Physics, University of Ioannina and Hellenic Institute of Nuclear Physics, Ioannina, Greece}
\author{L. Pandola}
\affiliation{INFN - Laboratori Nazionali del Sud, Catania, Italy}
\author{V. Soukeras}
\affiliation{Dipartimento di Fisica e Astronomia "Ettore Majorana", Universit\`a di Catania, Catania, Italy}
\affiliation{INFN - Laboratori Nazionali del Sud, Catania, Italy}
\author{G. Souliotis}
\affiliation{Department of Chemistry, University of Athens and Hellenic Institute of Nuclear Physics, Athens, Greece}
\author{A. Spatafora}
\affiliation{Dipartimento di Fisica e Astronomia "Ettore Majorana", Universit\`a di Catania, Catania, Italy}
\affiliation{INFN - Laboratori Nazionali del Sud, Catania, Italy}
\author{D. Torresi}
\affiliation{INFN - Laboratori Nazionali del Sud, Catania, Italy}
\author{A. Yildirim}
\affiliation{Department of Physics, Akdeniz Universitesi, Antalya, Turkey}
\author{V. A. B. Zagatto}
\affiliation{Instituto de F\'isica, Universidade Federal Fluminense, Niter\'oi, Brazil}
\collaboration{for the NUMEN collaboration}
\noaffiliation
\date{\today}
\begin{abstract}
%Single-nucleon transfer reactions are prominent spectroscopic tools characterized by high degree of selectivity in populating single-particle configurations in the final nuclear states. Furthermore, these reactions constitute one of the possible pathways of the complete double charge exchange mechanism. In principle, the measured double charge exchange cross-section is influenced by multi-step processes like successive single-nucleon transfer. Hence, the analysis of one-nucleon transfer cross-sections conveys all the necessary information to constrain the predicted cross-sections for such competitive processes.

The present article reports new data on the $^{48}$Ti($^{18}$O,$^{17}$O)$^{49}$Ti reaction at 275 MeV incident energy as part of the systematic research pursued within the NUMEN project. Supplementary measurements of the same reaction on $^{16}$O and $^{27}$Al targets were also performed in order to estimate the background arising from the use of a composite target (TiO$_{2}$ + $^{27}$Al). These data were analyzed under the same theoretical framework as those obtained with the titanium target in order to reinforce the conclusions of our analysis. Differential cross-section angular distribution measurements for the $^{17}$O$^{8+}$ ejectiles were performed in a wide angular range by using the MAGNEX large acceptance magnetic spectrometer. The experimental results were analyzed within the distorted-wave and coupled-channels Born Approximation frameworks. The optical potentials at the entrance and exit channels were calculated in a double folding approach adopting the S\~ao Paulo potential, and the spectroscopic amplitudes for the projectile and target overlaps were obtained from large-scale shell model calculations. The differential cross-sections are well-described by the theoretical calculations, where a weak coupling to collective excitations of projectile and target is inferred. The sensitivity of transfer cross-sections on different model spaces adopted in nuclear structure calculations, is also discussed.\par
\end{abstract}
\maketitle
\vspace{0.2cm}
\vspace{0.3cm}
%*********************************************
%
%
\section{Introduction}
%
%
%P. J. A. Buttle Nuclear Physics A176 (1971) 299-320
%High yields are observed for residual states presenting a high degree of overlap with the states at the entrance channel.
Over the past few years, studies for the neutrinoless double $\beta$ (0$\nu$$\beta$$\beta$) decay continue with undiminished interest, since it is considered the best probe of neutrino nature \cite{iachello,vergados,shimizu,ejiri,lenske,liang,capp_review_2023}. Moreover, if the 0$\nu$$\beta$$\beta$ decay is to be observed, the neutrino absolute mass scale could be extracted from the measured half-life \cite{ejiri,engel}. However, the latter is hampered by our limited knowledge of the nuclear matrix elements (NMEs) for this exotic process which, to date, their values are model dependent and susceptible to large uncertainties leading to vague conclusion for the neutrino absolute mass scale \cite{matrix_uncert,agostini}. High quality experimental data from single charge exchange \cite{sce_guess} or light-ion induced transfer \cite{schiffer1,schiffer2,schiffer3} reactions have been invoked to constrain the NMEs theories, but the ambiguities in the nuclear structure models are still too large. In this respect, more experimental constraints using different probes are highly desirable.\par 
Recently, a seminal experimental campaign was initiated at Istituto Nazionale di Fisica Nucleare - Laboratori Nazionali del Sud (INFN-LNS) in Catania within the NUMEN (NUclear Matrix Elements for Neutrinoless double $\beta$ decay) and NURE (NUclear REactions for neutrinoless double $\beta$ decay) \cite{nure} projects. NUMEN proposes an innovative experimental approach \cite{numen_cappu,numen_tdr} aiming at accessing information on the NMEs of 0$\nu$$\beta$$\beta$ decay by means of the heavy-ion induced double charge exchange (DCE) reactions on various $\beta$$\beta$ decay candidate targets \cite{capp_review_2023}. This may be achieved by collecting precise information on the NMEs of the DCE reaction \cite{lenske} which were recently proved to be correlated with the NMEs of the 0$\nu$$\beta$$\beta$ decay \cite{shimizu,santopinto}. However, in order to obtain meaningful information of the NMEs of DCE reactions, a comprehensive description of the complete DCE reaction mechanism is imperative.\par 
The direct meson exchange DCE reaction is one of the possible pathways of the complete DCE mechanism. The same final nuclear states can be, in principle, populated through two successive single charge exchange (SCE) reactions \cite{bellone}, through multi-nucleon transfer reactions \cite{ferreira_multi_nucleon} or a combination between them \cite{lay}. A recent theoretical work for the $^{20}$Ne+$^{116}$Cd reaction at 306 MeV \cite{ferreira_multi_nucleon} pointed to a small contribution of multi-nucleon transfer to the DCE cross-section, provided the appropriate experimental constraints \cite{carbone_2p,burrello} on the reaction models. Moreover, multi-nucleon transfer reactions are driven from mean-field dynamics and are susceptible to the kinematical matching conditions of Brink \cite{brink}. However, the contribution of all competitive processes to the measured DCE cross-section should not be considered ``a priori'' negligible, instead global studies of the DCE reactions \cite{capp_40ca,calabrese_dce,soukeras} together with all the available reaction channels are necessary \cite{burrello,cavallaro_frontiers,spatafora_multi}.\par
Over the past decades, single-nucleon transfer reactions induced by light ions have been established as a prominent tool of nuclear spectroscopy \cite{schiffer1,stiliaris,korsheninnikov,skaza,cavallaro_li10}. In early 50's, Butler demonstrated that the angular distribution pattern in (d,p) reactions is strongly dependent on the transferred angular momentum, thus providing information on the spin and parity of the final state nuclear wave function \cite{butler}. Additional information may be obtained from the analysis of the cross-section angular distributions. Single-nucleon transfer reactions favor transitions to final states with high degree of overlap with the states at the entrance channel allowing the determination of the spectroscopic factors. Further on, with the advent of radioactive beams these reactions have been employed to study the structure of exotic nuclei far from the valley of stability \cite{thomas,nature_jones,margerin}.\par
On the other side, heavy-ion induced transfer reactions have been less studied than their light ion counterparts. The origin of this shortage is likely twofold: the late dawn of powerful accelerators and high resolution detection systems for heavy ions suitable for spectroscopic studies and the supposed higher degree of complexity in heavy-ion dynamics. The latter consists in the strong dependence on the reaction cross-sections on Q-value and angular momentum matching conditions \cite{siemens,broglia,morrison} and the presence of strong absorption phenomena which should be effectively taken into account in the theoretical model of the reaction. An additional complication may come from the effect of inelastic excitations prior to and/or after the nucleon transfer process \cite{sinclair,tamura,linares_1n,keeley_ccba}. However, a substantial progress has been made in the field, demonstrating the usefulness of heavy-ion induced reactions in studies for nuclear structure and the reaction mechanism \cite{birnbaum,oertzen,scott,ursula,anyas,oertzen2,guidry,szilner,cappu_nature,cardozo}.\par  
The analysis of heavy-ion induced transfer reactions is usually performed in full quantum mechanical approach within the distorted-wave Born approximation (DWBA) formalism \cite{tobocman,satchler_dwba,timofeyuk_dwba}. When inelastic excitations prior or after transfer become important, the coupled-channels Born approximation (CCBA) model could be employed \cite{ccba}. In either theoretical framework, the calculation depends strongly on the optical potentials (OPs) and the overlap functions. The former is used to build the distorted waves at the entrance and exit channels which describe the direct elastic scattering between the interacting nuclei and also effectively account for absorption phenomena towards other non-elastic processes through an imaginary term. The OPs can be determined in phenomenological or double-folding frameworks, but in both cases their validity should be checked against elastic scattering data. On the other hand, the structure of the involved nuclei is inherent in the overlap functions. These quantities are approximated as solutions of a mean-field potential with fixed geometry weighted by the corresponding spectroscopic amplitude derived from a nuclear structure model. The advent of powerful computers facilitates the processing of shell model calculations with large model spaces for the determination of the spectroscopic amplitudes. However, certain limitations do exist like the case of nuclei well-beyond closed shells, where a reliable truncation of the full Hilbert space becomes imperative (e.g. \cite{yoshinaga,corrragio_shell_model_trunc}). In this sense, transfer reactions can be used to test the validity of nuclear shell model and improve our understanding of the mean-field description of nuclei.\par
On the above grounds, the description of heavy-ion induced transfer reactions is nowadays reasonably under control \cite{linares_1n,cappu_nature,montanari,caval,ermamatov,carbone2,ermamatov2,crema,zagatto,linares2}. During the past few years, a systematic study of one- and two- nucleon transfer reactions has been initiated by our group for the needs of the NUMEN project \cite{burrello,carbone_2p,spatafora_multi,sgouros_titan,calabrese,ciraldo,ciraldo_1p}. Transfer reactions can provide valuable information on the wave functions of nuclei involved in the $\beta$$\beta$ decay process and in conjunction with the information obtained from the study of the DCE reactions can provide experimental constraints on the nuclear structures models to be employed for the determination of the NMEs of 0$\nu$$\beta$$\beta$ decay.\par
%
%
%
%
%**************Figure 1***************************
\begin{figure}
\includegraphics[width=0.40\textwidth]{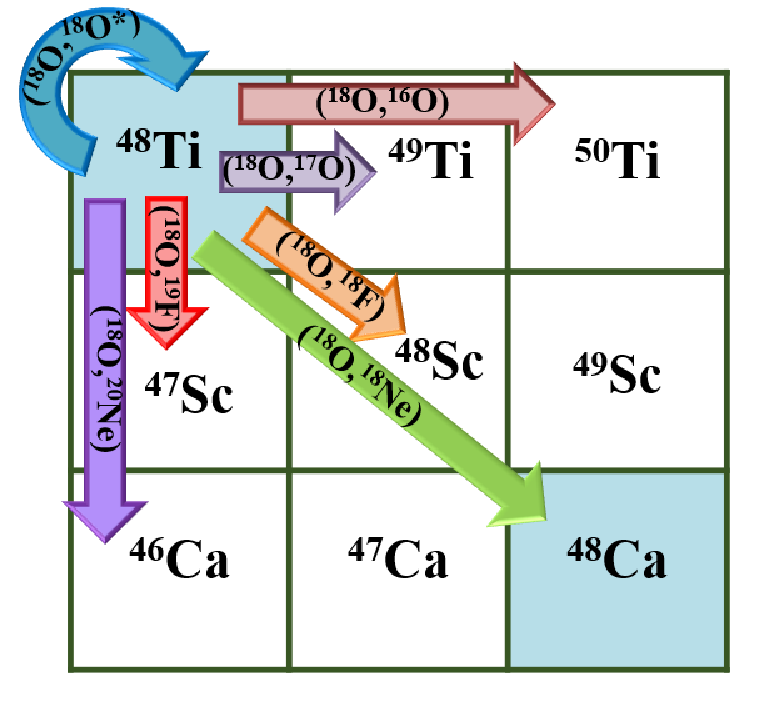}
\caption{Schematic representation of the nuclear reaction network measured in the $^{18}$O+ $^{48}$Ti collision. Arrows indicate the reaction paths connecting the initial and final partitions. The cyan curved arrow represents the elastic and inelastic scattering transitions.}
\end{figure}
Extending the aforementioned systematic, a global description in the $^{18}$O+$^{48}$Ti reaction at 275 MeV was performed under the NUMEN project by measuring all the available reaction channels. A schematic representation of the explored reaction network is illustrated in Fig.1. The analysis of the one-proton transfer reaction has been already completed \cite{sgouros_titan}, while the analysis of the elastic channel will be the subject of a forthcoming publication \cite{brischetto_elast}. The $^{48}$Ti was used as target since is the daughter nucleus of $^{48}$Ca in the 0$\nu$$\beta$$\beta$ decay process \cite{48ca_double_beta_decay,48ca_beta_beta,abinitio_48ca}. Therefore, all possible information about its structure is essential for the determination of the relevant NMEs. The present article is dedicated to the study of single-neutron transfer reaction channel. The angular distribution of the measured cross-sections are analyzed within the DWBA formalism for validating the optical potential at the entrance channel which was deduced in a parallel analysis of the elastic scattering data \cite{brischetto_elast} as well as to test the validity of adopted the spectroscopic amplitudes derived from large-scale shell model calculations. The role of inelastic excitations prior to transfer process was also investigated by means of CCBA calculations. Further on, supplementary measurements of the same reaction on $^{16}$O and $^{27}$Al targets were performed in order to estimate the background arising from the different target components (TiO$_2$+$^{27}$Al) as well as to reinforce the conclusions of our analysis. Additionally, this work highlights the selectivity of transfer reactions on the model space used in the nuclear structure calculations and complements the work by Ciraldo et al. \cite{ciraldo}, where the sensitivity of transfer reactions on the adopted nuclear structure model was inferred.\par
In what follows, the experimental setup and data reduction are reported in Sections II and III, respectively, whereas in Section IV the description of our theoretical approach for the calculation of the one-neutron transfer cross-sections is presented. The results of the analysis are discussed in Section V and finally, some concluding remarks are given in Section VI.
\section{Experimental setup}
The experiment was conducted at the MAGNEX facility of INFN-LNS laboratoty in Catania, Italy. A fully stripped $^{18}$O ion beam, accelerated at 275 MeV by the K800 Superconducting Cyclotron, impinged on a (510 $\pm$ 26) $\mu$g/cm$^2$ TiO$_{2}$ target which was evaporated onto a thin aluminum foil (216 $\pm$ 11 $\mu$g/cm$^2$). Supplementary measurements, under the same experimental conditions, were repeated using a $^{27}$Al target (226 $\pm$ 11 $\mu$g/cm$^2$) as well as a WO$_{3}$ one (284 $\pm$ 14 $\mu$g/cm$^2$) with an aluminum backing for estimating the background contribution. The beam charge was integrated by means of a Faraday cup placed 150 mm downstream the target position. An electron suppressor ring polarized at -200V was placed at the entrance of the Faraday cup to mitigate the error in the beam current measurement due to the escape of secondary produced electrons.\par
%
%
%
%
%**************Figure 2***************************
\begin{figure}
\includegraphics[width=0.50\textwidth]{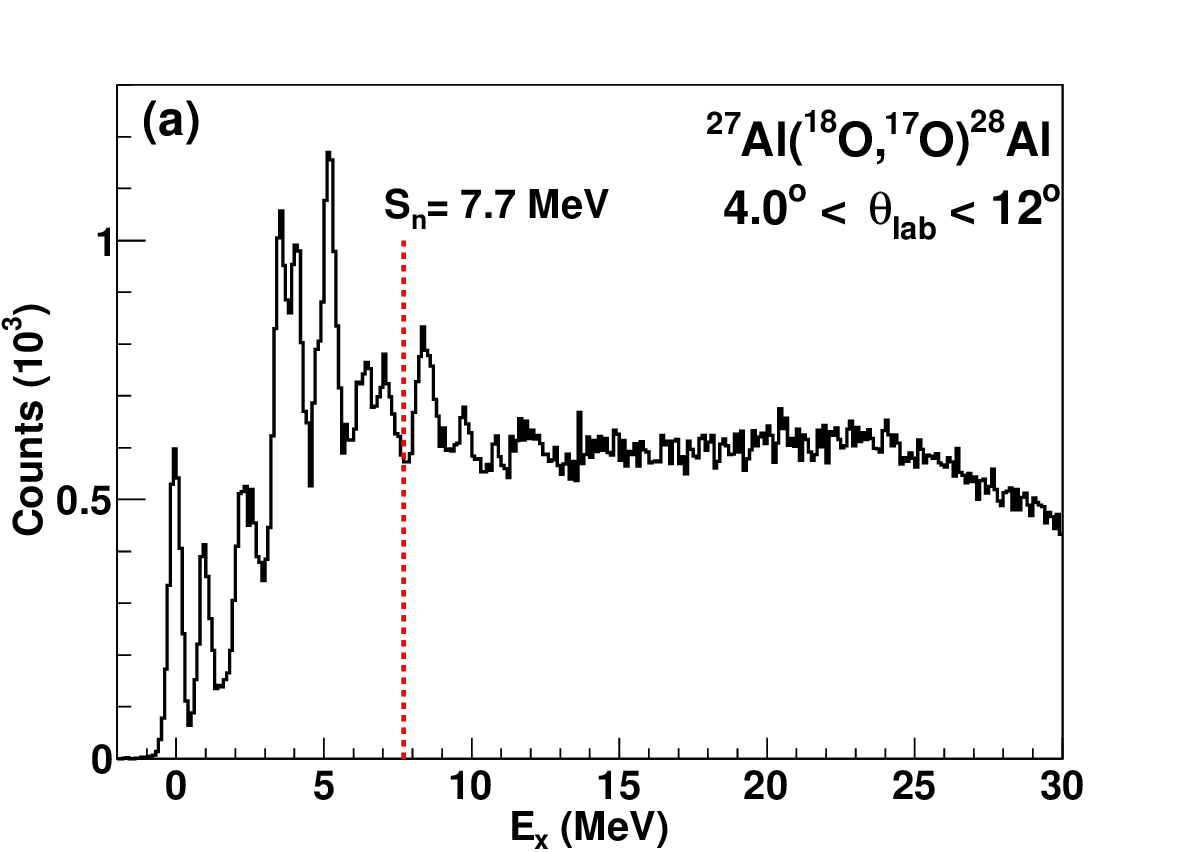}
\vspace{0.1cm}
\includegraphics[width=0.50\textwidth]{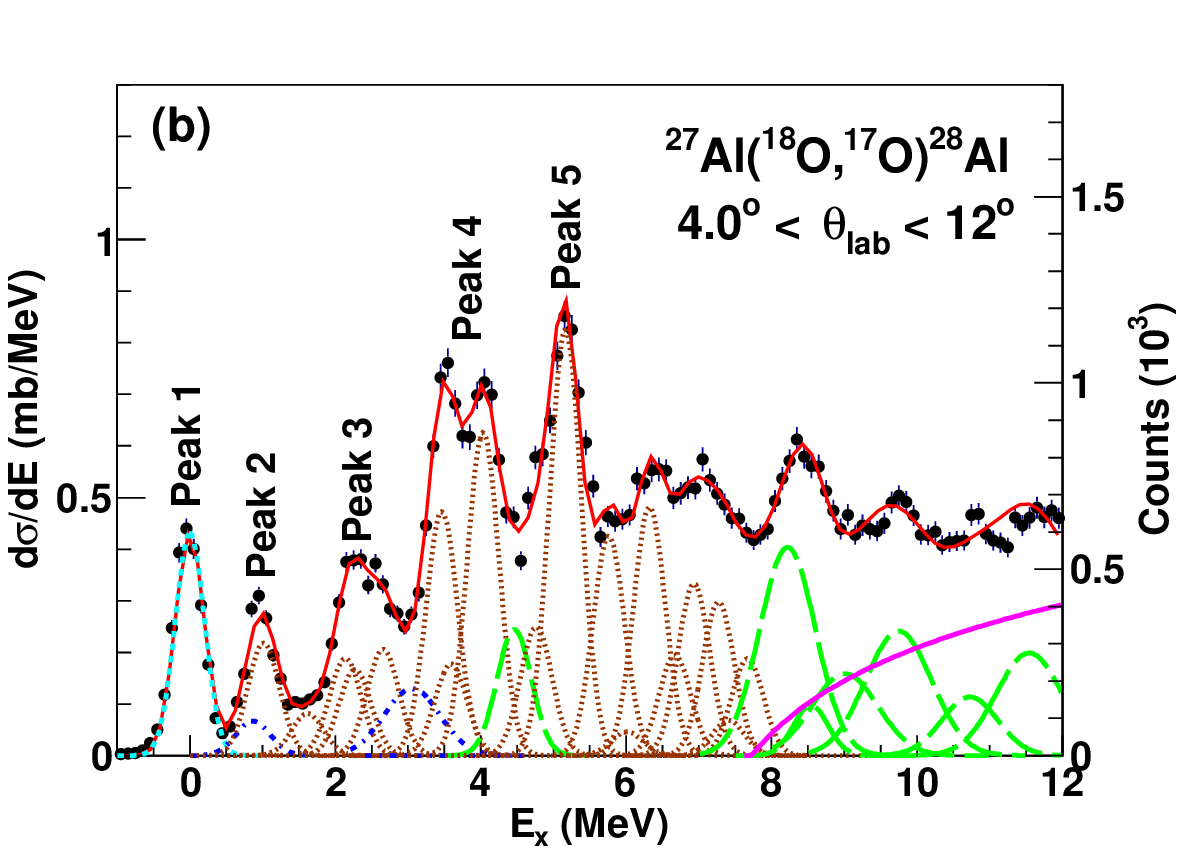}
\caption{(a) Reconstructed excitation energy spectrum for the $^{27}$Al($^{18}$O,$^{17}$O)$^{28}$Al one-neutron transfer reaction measured at 275 MeV. The neutron separation threshold, S$_{n}$, of $^{28}$Al nucleus is denoted with the dashed red line. (b) Energy distribution for the $^{27}$Al($^{18}$O,$^{17}$O)$^{28}$Al reaction. The experimental data are compared to the result of a multi-fit analysis, where each nuclear state is represented by a Gaussian form factor. The cyan dashed curve represents the g.s. to g.s. transition, the blue dotted-dashed curves represent the excited states of $^{17}$O nucleus, the brown dotted curves correspond to excited states of $^{28}$Al nucleus with the $^{17}$O one being in the g.s. and the long-dashed green curves represent transitions where both the $^{17}$O and $^{28}$Al nuclei are excited. The continuum background above the single-neutron emission threshold of $^{28}$Al is indicated with the solid magenta line. The solid red line corresponds to the sum of the individual Gaussian curves.}
\end{figure}
The reaction ejectiles were momentum analyzed by the MAGNEX large acceptance magnetic spectrometer \cite{magnex} whose optical axis was set at $\theta_{opt}$= \ang{9}, thus subtending an angular range between \ang{3} and \ang{15} in the laboratory reference frame. The large acceptance in momentum of the spectrometer (-14\%,+10.3\% with respect to the optical axis) allowed the simultaneous transport at the MAGNEX focal plane of the $^{17}$O$^{8+}$ and  $^{19}$F$^{9+}$ ions coming from the one-neutron and one-proton transfer reactions, respectively. The solid angle, delimited by four slits at the entrance of the spectrometer, was reduced to 14 msr for preventing the deterioration of the Focal Plane Detector (FPD) \cite{fpd,fpd2} response from the high ion rate. The FPD is a hybrid detector comprised a proportional drift chamber, served as energy loss ($\Delta$E) detector and as tracker, followed by a wall of 60 silicon detectors for measuring the ions residual energy (E$_{r}$). Using the information provided by the FPD, the particle identification (PID) was performed. Details pertinent to this work are provided herewith, while a description of the PID technique has been discussed in details elsewhere \cite{pid}. The discrimination of the various ion species based on their atomic number was performed adopting the $\Delta$E-E$_{r}$ technique, while for a given ion the different isotopes were discriminated following a technique based on the correlation between the ions kinetic energy and the measured position along the dispersive direction (horizontal in the case of MAGNEX). Examples of the PID technique can be found in Refs.\cite{sgouros_titan,pid,nim_16o_27al_elast,fonseca,calabrese_nim,sgouros_sif}.
\section{Data reduction}
Having identified the $^{17}$O$^{8+}$ events, a software ray reconstruction was performed for each of the available data sets. This procedure allows one to obtain the momentum vector of the ions at the target position \cite{momentum_vector} from the measured coordinates at the reference frame of the FPD.
\subsection{The $^{27}$Al($^{18}$O,$^{17}$O)$^{28}$Al reaction}
Following the same strategy as the one adopted in the case of the one-proton transfer reaction channel \cite{sgouros_titan}, we present the results for the $^{27}$Al($^{18}$O,$^{17}$O)$^{28}$Al reaction considering that the aluminum backing constitutes a substantial source of background in the measurements with the TiO$_2$ and WO$_3$ targets. The excitation energy, E$_{x}$, was determined as:
%
%
%**************Equation 1***************************
\begin{equation}
E_{x} = Q_{0}-Q,
\end{equation} 
where \textit{Q$_{0}$} is the ground state (g.s.) to g.s. Q-value for the $^{27}$Al($^{18}$O,$^{17}$O)$^{28}$Al reaction calculated from the mass imbalance at the entrance and exit channels and \textit{Q} is the reaction Q-value calculated adopting the missing mass method \cite{magnex} based on relativistic two-body kinematics. Given the large acceptance in momentum of the spectrometer, the excitation energy spectrum for the $^{27}$Al($^{18}$O,$^{17}$O)$^{28}$Al reaction was measured in a wide energy range and is presented in Fig. 2a. The achieved energy resolution was $\approx$ 400 keV full width at half maximum (FWHM). After correcting the experimental yields for the overall efficiency of the spectrometer \cite{effic} and by taking into account the solid angle and the beam flux of the measurement, the absolute energy differential cross-sections were deduced and are presented in Fig. 2b. The spectrum is characterized by a sharp increase in the cross-section between 3.5 and 6 MeV, which is attributed to transitions to the negative parity states of $^{28}$Al at 3.46, 3.59, 4.03 and 5.17 MeV owing to an appreciable single-particle strength. This is compatible with what has been previously observed in (d,p) reaction experiments \cite{alum28_carola}, while it is further corroborated by our theoretical calculations which are presented in the following section. Considering that the $^{17}$O nucleus has only 3 bound excited states, the shape of the energy spectrum is mainly determined by the level density of the $^{28}$Al. A multi-fit analysis performed by using as inputs the known states of $^{17}$O and $^{28}$Al nuclei, reproduces the overall shape of the energy distribution adequately-well. It should be underlined that the individual contribution of each transition could not be extracted from the fit, but the overall contribution of the various unresolved states. To this extent, the experimental yields corresponding to five discrete groups of states were deduced from the fits considering a variable angular step (between \ang{0.5} and \ang{2}), depending on the available yield. The resulted differential cross-section angular distributions are presented in Fig. 3. The error in the angular distribution data is dominated from the statistical uncertainty and, to a lesser extent, from the uncertainty in the determination of the solid angle. An overall error of about 10 \% due to the uncertainty in the target thickness and the integrated value of the beam charge, common to all the data points, is not included in the error bars. The shape of the angular distribution data is characterized by a sharp fall-off in the differential cross-section ascribed to strong absorption phenomena manifested after the grazing angle (\(\theta\)$_{gr}^{c.m}$= \ang{5.6}).\par
%
%
%
%**************Figure 3*************************** 
\begin{figure*}
\centering
\includegraphics[width=0.80\textwidth]{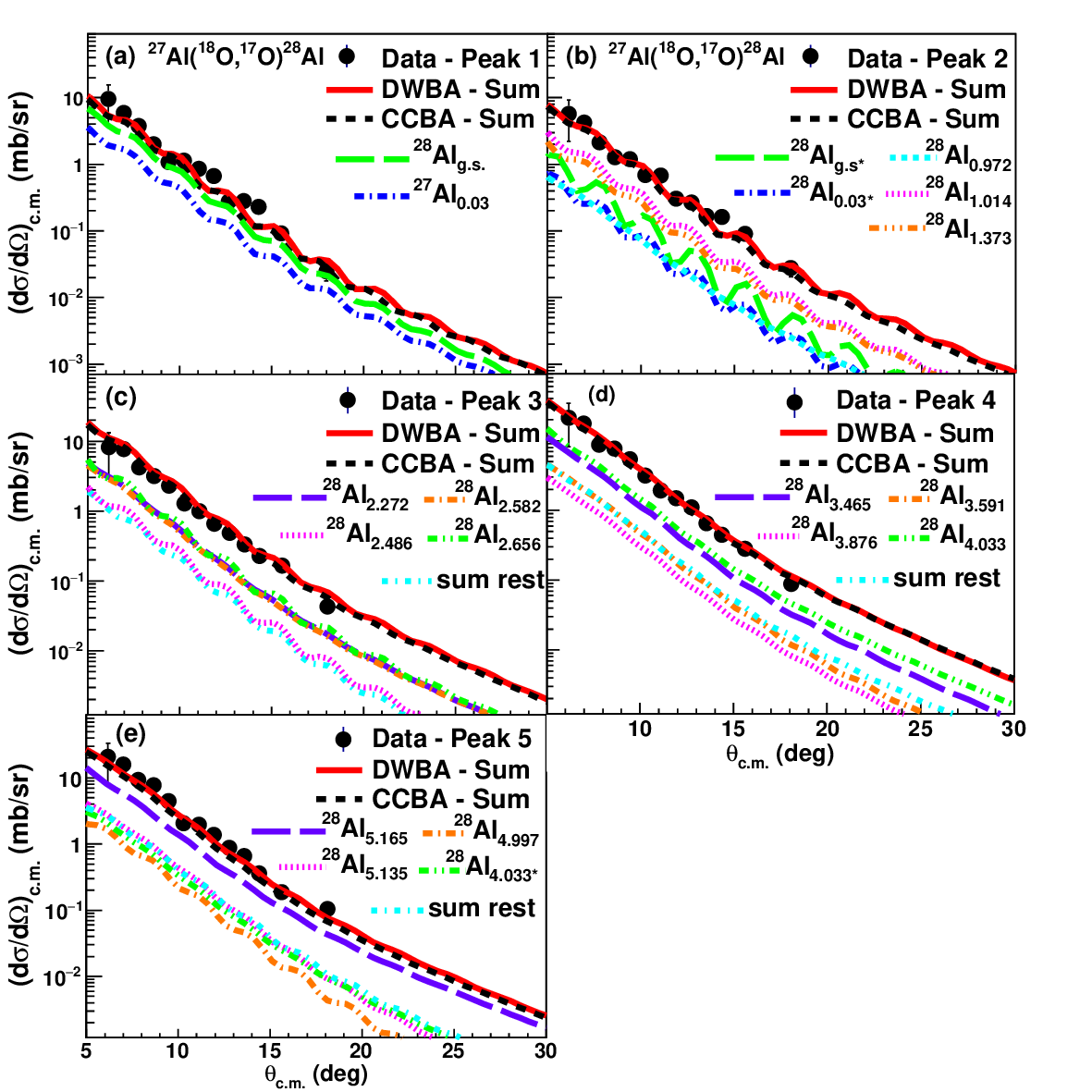}
\caption{Comparison between experimental data and theoretical predictions for the $^{27}$Al($^{18}$O,$^{17}$O)$^{28}$Al one-neutron transfer reaction at 275 MeV. The experimental data which are indicated by the black points correspond to five excitation energy regions in Fig. 2b labeled as (a) Peak 1, (b) Peak 2, (c) Peak 3, (d) Peak 4 and (e) Peak 5. Theoretical angular distribution cross-sections for the transitions to the involved states of the ejectile and the residual nuclei were calculated within the DWBA framework and are presented with the colored curves. In the legend, each curve is labeled by the corresponding excitation energy of $^{28}$Al for transitions to the $^{17}$O$_{g.s.}$ and by an asterisk in case where $^{17}$O is excited to the \(\frac{1}{2}\)$_{1}^{+}$ (0.871 MeV) state. In panels (c)-(e), for reasons of clarity only the four stronger transitions are shown, while the contribution for the rest of transitions is indicated with the cyan dotted-dashed curve under the notation "sum rest". In all cases, the sum of all transitions is illustrated by the red solid line. The overall result of a CCBA calculation considering the same final states as the DWBA one is illustrated with the dashed black curve.}
\end{figure*}
\subsection{The $^{16}$O($^{18}$O,$^{17}$O)$^{17}$O reaction}
The excitation energy spectrum of the $^{16}$O($^{18}$O,$^{17}$O)$^{17}$O one-neutron transfer reaction is presented in Fig. 4a. The distinct structures located at E$_{x}$ $\approx$ -4.6 and -3.7 MeV were identified as one-neutron transfer events originated from the reaction of $^{18}$O beam with the aluminum backing of the target. Thus, using the previously analysed data collected in the measurement with a self-supporting aluminum target, the energy distribution of the contaminant events, appropriately normalized, was subtracted from the total spectrum. The normalization factor was chosen such as the ratio of the integral for the peaks at E$_{x}$ $\approx$ -4.6 and -3.7 in the two data sets to be $\approx$ 1. The resulted normalization factor was found to be the same as the one deduced in a parallel analysis for the one-proton transfer reaction on the same target \cite{sgouros_titan} which was measured under the same experimental conditions. A very small contribution coming from the one-neutron transfer reaction with the tungsten component of the target was further subtracted assuming a uniform distribution throughout the whole energy range. This distribution was normalized to a small remnant of events at E$_{x}$$<$-7 MeV, a region free of any contamination due to the reaction with the aluminum backing. After subtracting the contaminant events, the excitation energy spectrum corresponding to the $^{16}$O($^{18}$O,$^{17}$O)$^{17}$O reaction was deduced and it is presented in Fig. 4a. Subsequently, by correcting the experiment yields for the efficiency of the spectrometer and by taking into account the solid angle and the beam flux of the measurement, absolute energy and angle differential cross-sections were determined and are presented in Figs. 4b and 5, respectively. The error in the differential cross-sections includes the contribution from statistical, solid angle and background subtraction uncertainties. In this case, since the ejectile and recoil nuclei are identical the energy profile of the reaction contains, in general, contributions from transitions to excited states of both nuclei. The contribution of each transition to the measured cross-sections could be disentangled via theoretical calculations.\par
%The fact that two independent analyses suggest the same normalization factor speaks for the consistency of the data analysis.
%
%
%
%
%**************Figure 4***************************
\begin{figure}[!ht]
\begin{center}
\includegraphics[width=0.50\textwidth]{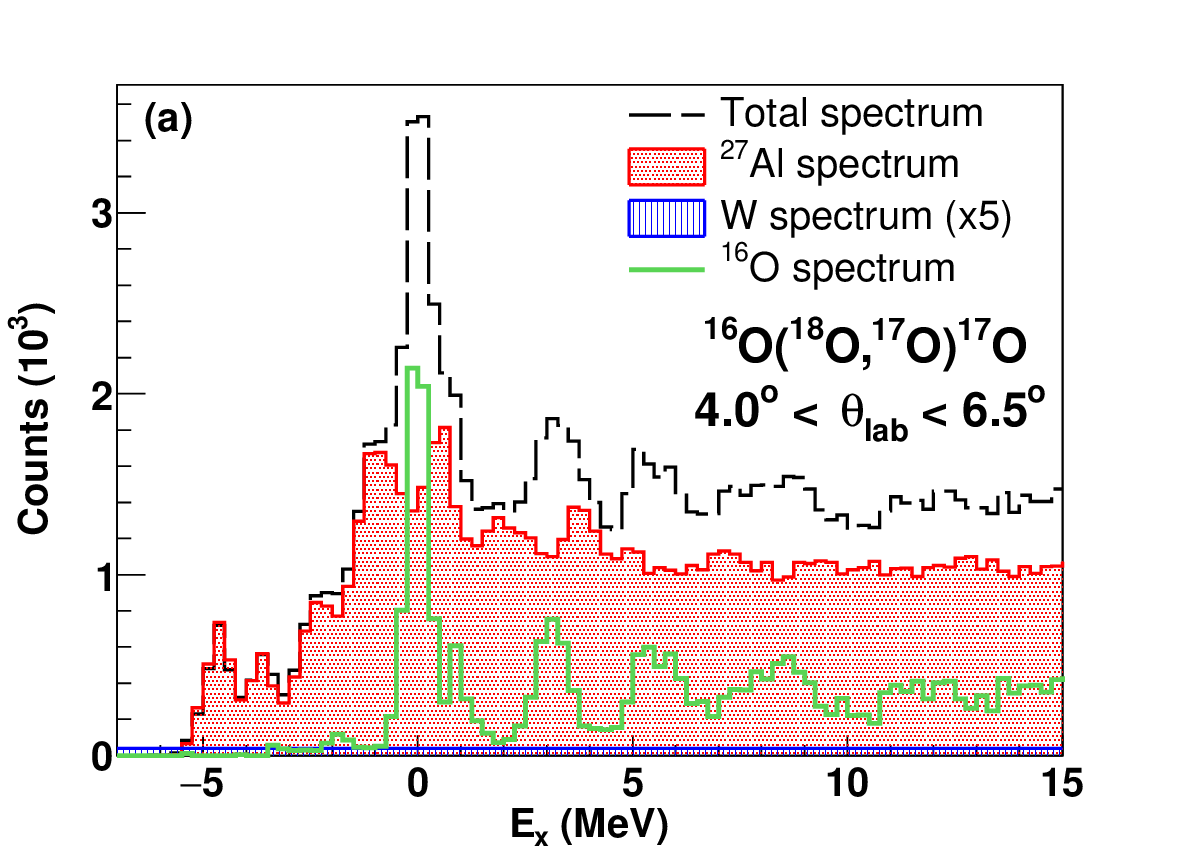}
\includegraphics[width=0.50\textwidth]{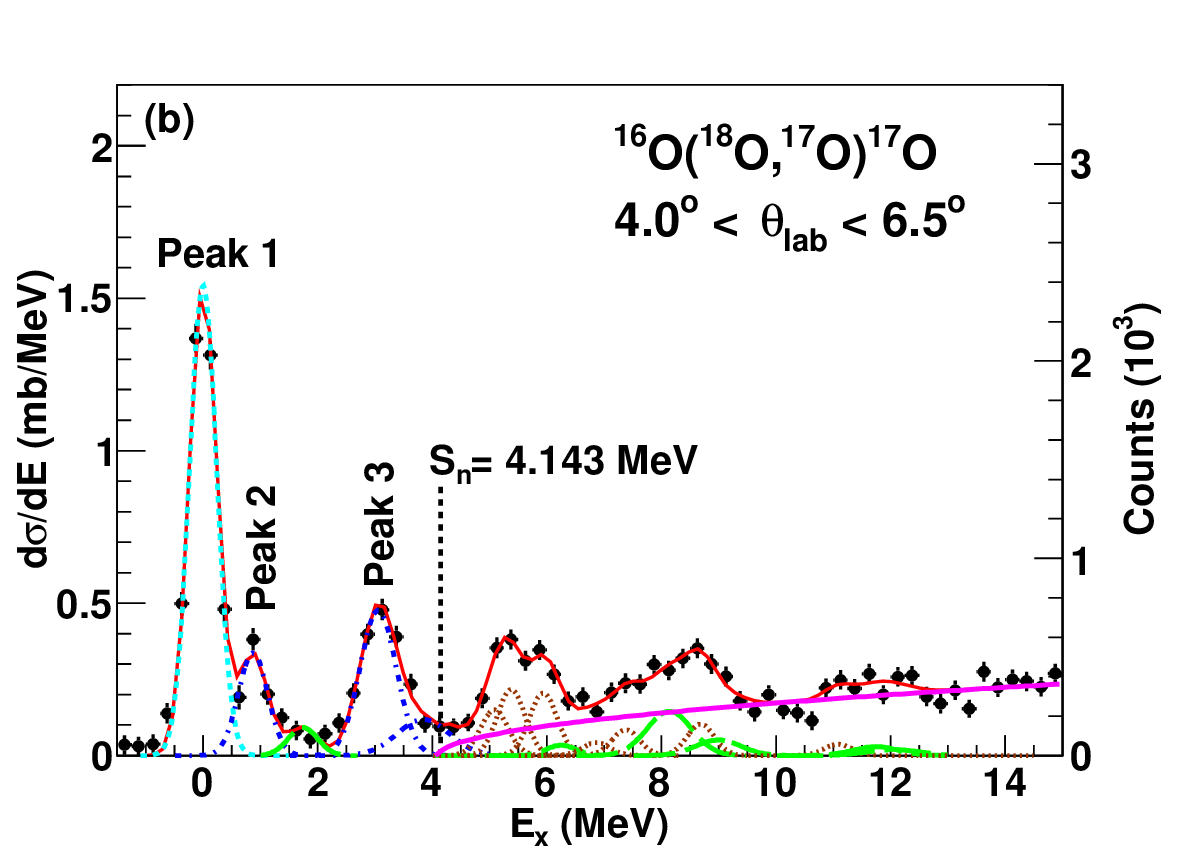}
\caption{(a) Decomposition of the excitation energy spectrum measured at 275 MeV using a WO$_{3}$ + $^{27}$Al target. The total spectrum is depicted with the dashed black line. The red-filled area represents the normalized background spectrum originated from the reaction of the beam with the aluminum backing, while the blue-hatched area corresponds to the background spectrum coming from the tungsten component of the target. For reasons of clarity, the latter was scaled by a factor of 5. The spectrum represented by the solid green line corresponds to the energy profile of the $^{16}$O($^{18}$O,$^{17}$O)$^{17}$O after subtracting from the total spectrum the background events. (b) Energy distribution for the $^{16}$O($^{18}$O,$^{17}$O)$^{17}$O reaction. The experimental data are compared to the result of a multi-fit analysis where each nuclear state is represented by a Gaussian form factor. The line styles and colors used for the description of each curve are the same as those presented in Fig. 2b. The single-neutron separation threshold, S$_{n}$, of $^{17}$O nucleus is denoted with the dashed black line.}
\end{center}
\end{figure}
%
%
%

%**************Figure 5*************************** 
\begin{figure}
\begin{center}
\includegraphics[width=0.50\textwidth]{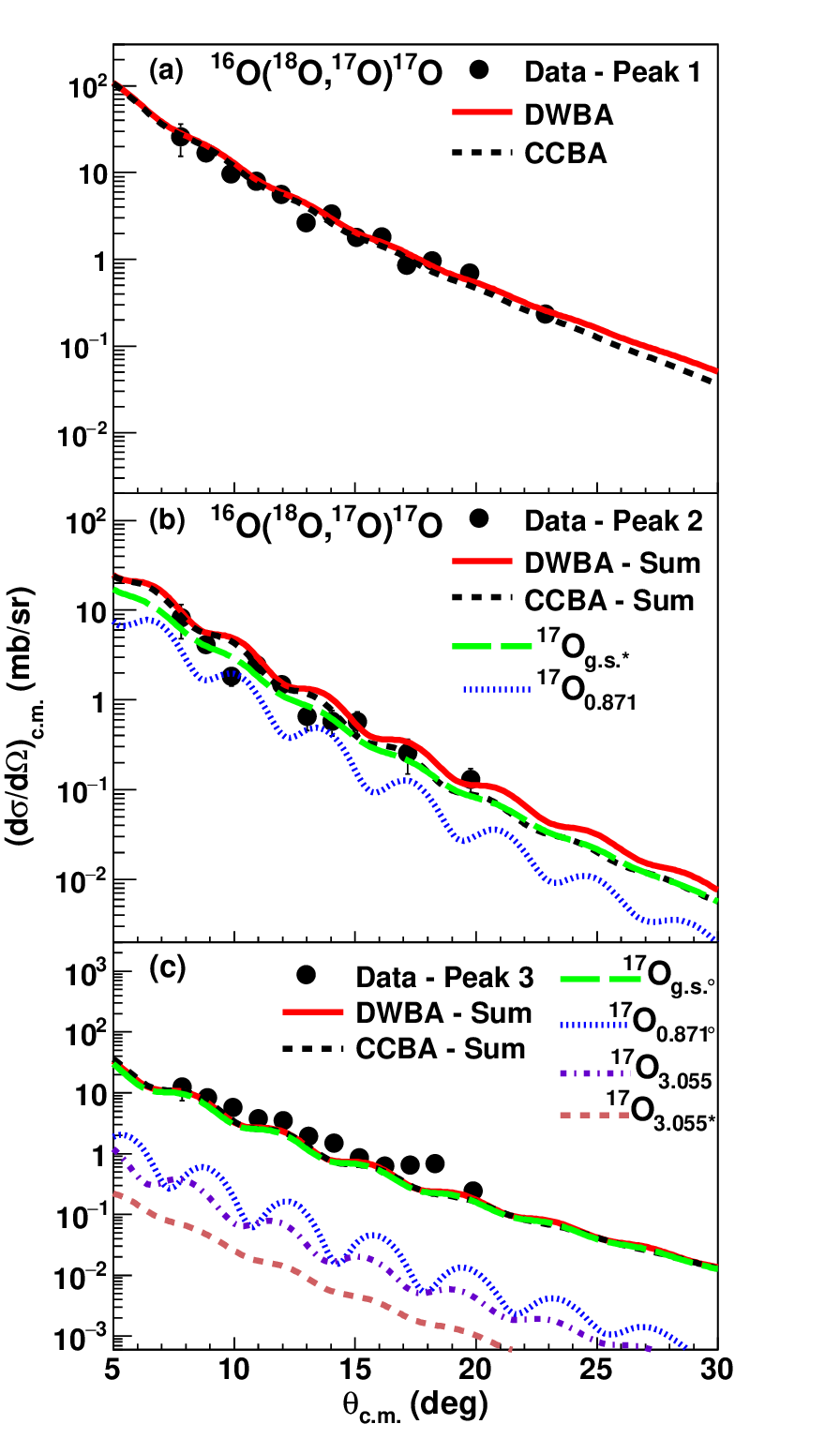}
\caption{Comparison between experimental data and theoretical predictions for the $^{16}$O($^{18}$O,$^{17}$O)$^{17}$O one-neutron transfer reaction at 275 MeV. The experimental data which are indicated by the black points correspond to three excitation energy regions in Fig. 4b labeled as (a) Peak 1, (b) Peak 2 and (c) Peak 3. Theoretical angular distribution cross-sections for the transitions to the involved states of the ejectile and the residual nuclei were calculated within the DWBA framework and are presented with the colored curves. In the legend, each curve is labeled by the corresponding excitation energy of the recoil nucleus for transitions to the g.s. of the ejectile and a symbol in case where $^{17}$O ejectile is excited. Curves marked with an asterisk and a circle refer to transitions to the \(\frac{1}{2}\)$_{1}^{+}$ (0.871 MeV) and \(\frac{1}{2}\)$_{1}^{-}$ (3.055 MeV) states of $^{17}$O, respectively. The sum of all transitions is illustrated by the red solid line. The overall result of a CCBA calculation considering the same final states as the DWBA one is illustrated with the dashed black curve.}
\end{center}
\end{figure}
\subsection{The $^{48}$Ti($^{18}$O,$^{17}$O)$^{49}$Ti reaction}
The excitation energy spectrum of the $^{48}$Ti($^{18}$O,$^{17}$O)$^{49}$Ti one-neutron transfer reaction is presented in Fig. 6a. The spectrum is contaminated by events generated by the reaction of the beam with the oxygen and aluminum components of the TiO$_{2}$+$^{27}$Al target. The yield due to such background sources was determined using the previously analysed data presented in Figs. 2a and 4a. The two background spectra were scaled using the same normalization factors as those deduced from the analysis of the one-proton transfer reaction data on the same target \cite{sgouros_titan}, since both reactions were measured under the same experimental runs. Subsequently, the background yields were subtracted and the excitation energy spectrum corresponding to the $^{18}$O + $^{48}$Ti reaction was deduced and is presented in Fig. 6a. Following the same procedure as in the previous two cases, the absolute energy differential cross-section as a function the excitation energy was determined and it is shown in Fig. 6b. The obtained energy resolution for this data set is $\approx$ 450 keV at FWHM.\par
 For the determination of the angular distribution cross-sections a different approach was followed for this data set. While in the case of the one-neutron transfer reaction with the aluminum and oxygen targets the counts were obtained from the area of the Gaussian functions, in this case, since the level density of the $^{49}$Ti nucleus is appreciably higher, the experimental yields were determined by integrating four regions of interest (ROI). In more details, up to the excitation energy of $\approx$ 2 MeV where the level density for $^{49}$Ti is rather low, tests adopting either the ROI or the Gaussian function approach yielded almost identical results. However, beyond such energy region it was found that the integrated yields are susceptible to the results of the fit. Therefore, in order to avoid a model-dependent analysis and thus increasing the systematic errors in the data reduction, it was decided to select four discrete ROIs as it is shown in Fig. 6b. The same prescription was also applied in the analysis of the one-neutron transfer reaction for the system $^{18}$O + $^{76}$Se \cite{ciraldo}, where the density of states in the $^{77}$Se nucleus is comparable to the one of $^{49}$Ti. The resulted angular distribution data are presented in Fig. 7.
%  
%
%
%
%**************Figure 6***************************
\begin{figure}[ht]
\begin{center}
\includegraphics[width=0.50\textwidth]{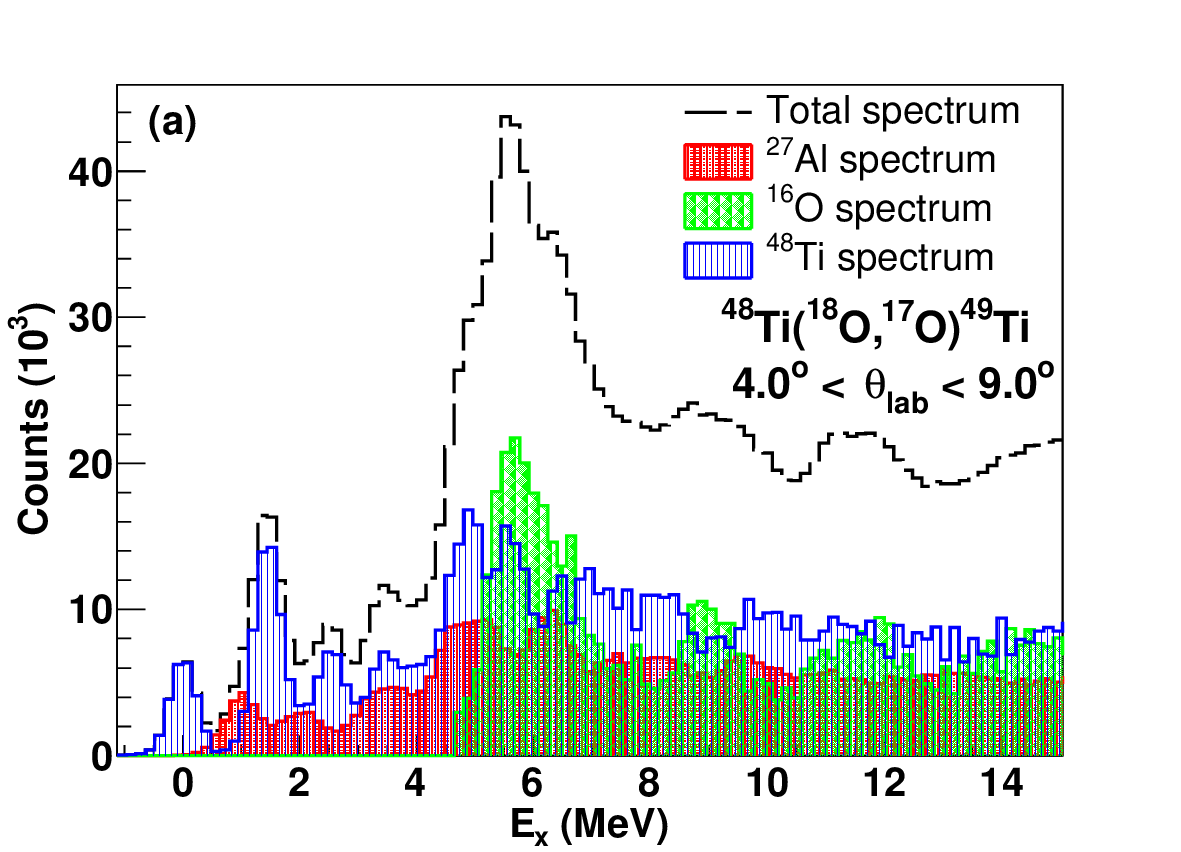}
\vspace{0.1cm}
\includegraphics[width=0.50\textwidth]{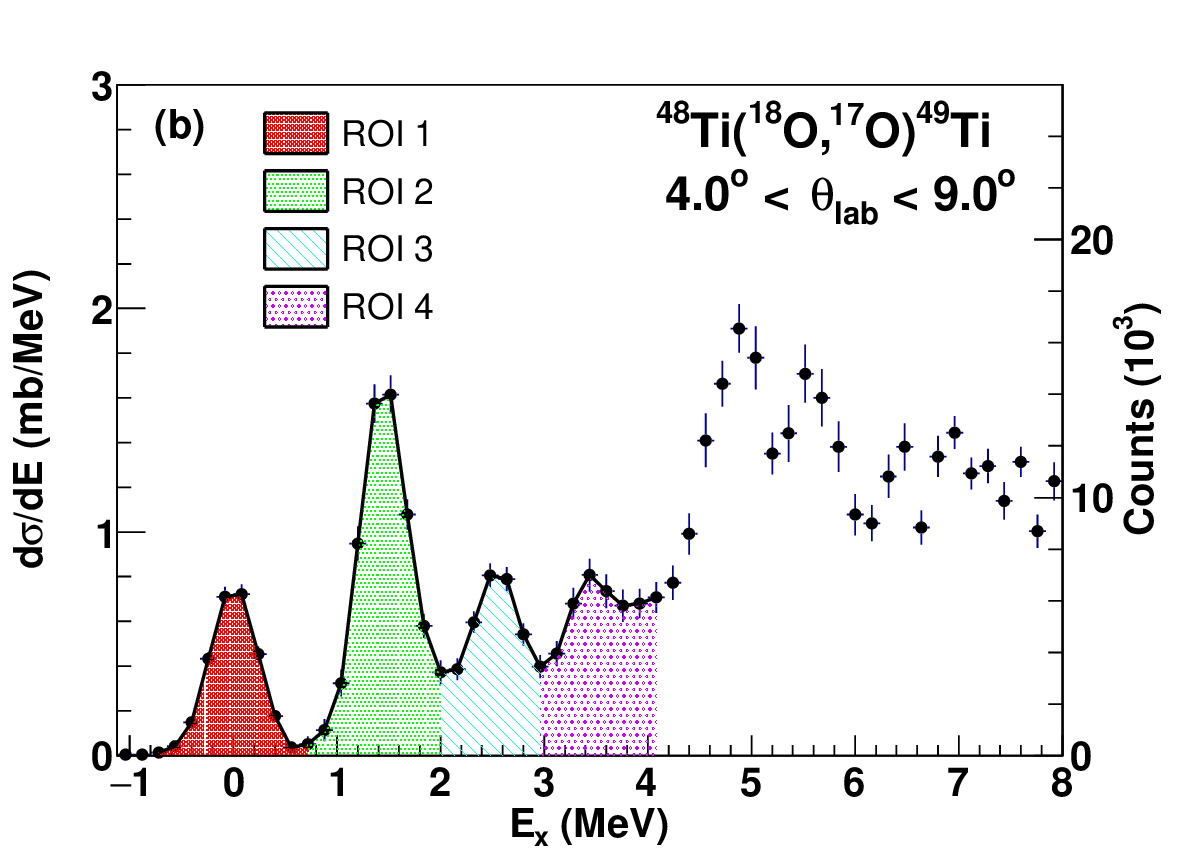}
\caption{(a) Decomposition of the excitation energy spectrum measured at 275 MeV using a TiO$_{2}$ + $^{27}$Al target. The total spectrum is depicted with the dashed black line. The red-dotted area represents the normalized background spectrum originated from the reaction of the beam with the aluminum backing, while the green-filled area corresponds to the background spectrum coming from the oxygen component of the target. The spectrum represented by the blue-hatched area corresponds to the energy profile of the $^{48}$Ti($^{18}$O,$^{17}$O)$^{49}$Ti, after subtracting from the total spectrum the background events. (b) Energy distribution for the $^{48}$Ti($^{18}$O,$^{17}$O)$^{49}$Ti reaction. The spectrum is divided in four regions of interest (ROI) which are indicated by the colored areas.}
\end{center}
\end{figure}
%
%

%**************Figure 7*************************** 
\begin{figure*}
\centering
\includegraphics[width=0.60\textwidth]{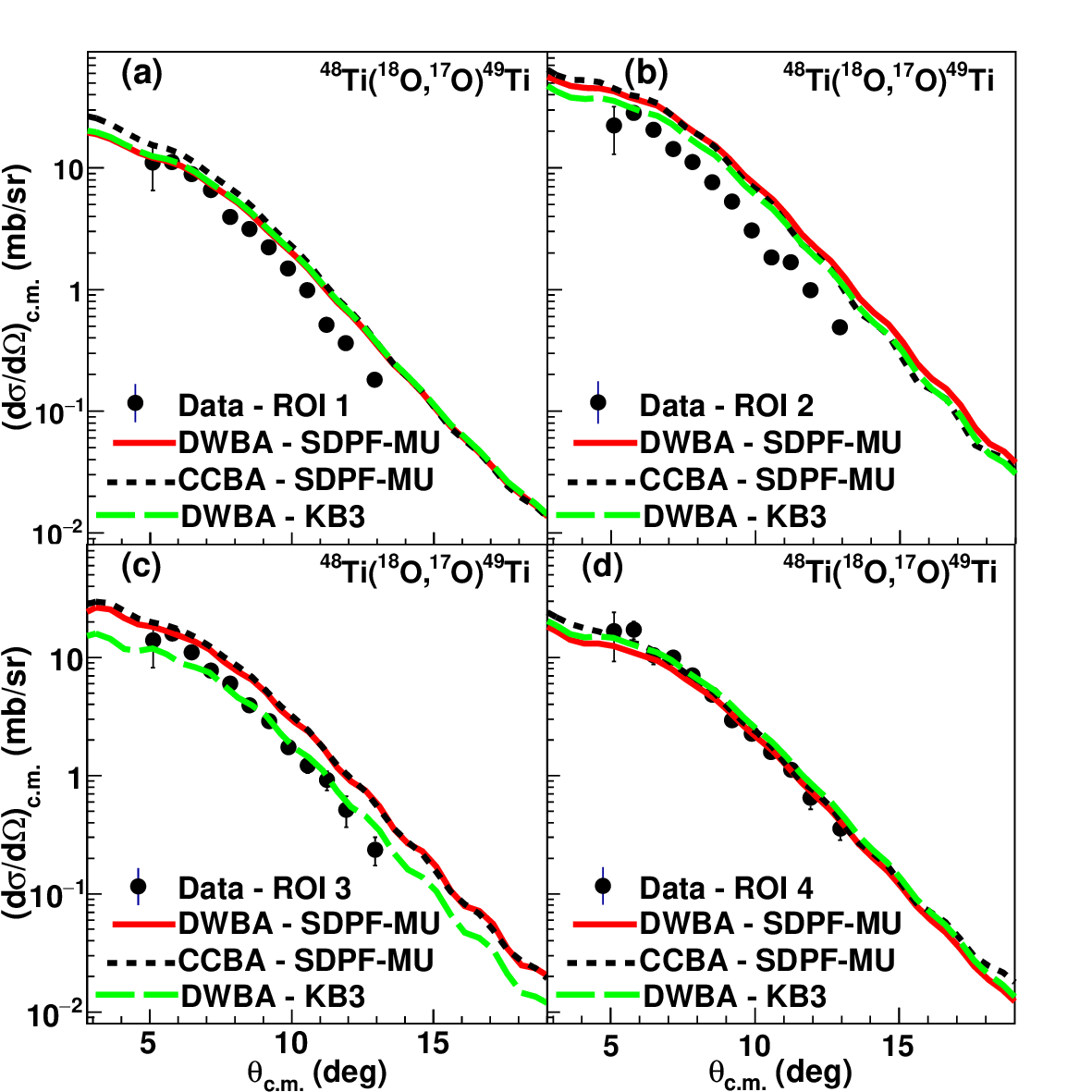}
\caption{Comparison between experimental data and theoretical predictions for the $^{48}$Ti($^{18}$O,$^{17}$O)$^{49}$Ti one-neutron transfer reaction at 275 MeV. The experimental data which are indicated by the black points correspond to four excitation energy regions in Fig. 6b labeled as (a) ROI 1, (b) ROI 2, (c) ROI 3 and (d) ROI 4. Theoretical angular distribution cross-sections were derived within the DWBA and CCBA frameworks and are presented with the colored curves. In the legend, each curve is marked with the name of the effective interaction which was employed in the nuclear structure calculations.}
\end{figure*}

\section{Theoretical analysis}
Theoretical cross-section angular distribution for the transfer reactions under study were calculated under the DWBA and the CCBA frameworks using the FRESCO code \cite{fresco}. The transfer operator was built employing the post-representation and incorporating full complex remnant terms. The distorted waves at the entrance and exit channels were generated using the double-folding S\~ao Paulo potential (SPP) \cite{spp1,spp2,spp3} for the description of the real and imaginary part of the optical potential, but adopting two different normalization coefficients for each part. The real part of the optical potential accounts for the refraction of the incident waves, while the imaginary part for all non-elastic processes \cite{hodgson}. The normalization coefficient of the real part, N$_{R}$, was equal to unity as typically adopted for heavy ion elastic scattering involving tightly bound nuclei above the Coulomb barrier \cite{satchler_love}. The value of the normalization factor of the imaginary part, N$_{I}$, may slightly vary depending on the adopted coupling scheme. In the DWBA calculations where no couplings to collective states of projectile and/or target are considered the value of N$_{I}$ was 0.78, whereas in the CCBA approach where couplings to the low-lying states of the projectile and the target were taken into account, this factor was further reduced to N$_{I}$= 0.60. This prescription has been successfully applied in the past for the description of various elastic scattering and single-nucleon transfer reaction data \cite{spatafora_multi,linares_1n,caval,sgouros_titan,calabrese,ciraldo,ciraldo_1p,pereira_elast,capp_elast,spatafora_elast,carbone_elast}. For the case of the $^{18}$O + $^{48}$Ti reaction, the optical potential at the entrance channel was checked against elastic and inelastic scattering data which were measured in the same experiment \cite{brischetto_elast}.\par
The bound state wave functions were calculated as single-particle solutions of the Schroedinger with a Woods-Saxon potential of certain geometry. For the $^{17}$O core the reduced radius and diffuseness of the potential were 1.26 and 0.70 fm, respectively, while the corresponding values for the heavier cores were 1.20 and 0.60 fm, respectively. This is a typical choice of parameters that has been adopted in many of our previous works for $^{18}$O induced transfer reactions on different target species \cite{spatafora_multi,sgouros_titan,ciraldo,ferreira}. For the $^{16}$O core which is a tightly bound doubly-magic nucleus, the adopted values for the reduced radius and diffuseness of the potential were 1.20 and 0.60 fm, respectively \cite{linares_1n}. In all cases the depth of the Woods-Saxon potential was varied such as to reproduce the experimental binding energies of the valence neutron. \par
In the CCBA calculations, couplings to the low-lying states of the projectile and target were included using the conventional rotational model, while no couplings to excited states of the nuclei at the exit partition were considered. In the rotational model, the deformation of the Coulomb potential is extracted from the intrinsic matrix element, while the strength of the nuclear deformation is calculated from the corresponding deformation length. Considering a transition from an initial state with spin 0$^{+}$, as in the present case, to a final state with spin \textit{J} under the action of an operator of multipolarity \textit{k}, the intrinsic matrix element, \textit{M$_{n}$}, is evaluated as:    
%
%
%**************Equation 2***************************
\begin{equation}
M_{n}(E_{k},0^{+}\rightarrow J) = \pm \sqrt{(2J+1)B(E_{k};0^{+}\rightarrow J)},
\end{equation} 
where \textit{B(E$_{k}$)} is the reduced transition probability. The relative sign of \textit{M$_{n}$(E$_{k}$)} is taken as that of the intrinsic quadrupole moment following the prescription of Ref. \cite{fresco}. The deformation length is related to the intrinsic matrix element through the following expression:
%
%
%**************Equation 3***************************
\begin{equation}
\delta_{k}= \frac{4\pi}{3Ze} \frac{M_{n}(E_{k},0^{+}\rightarrow J)}{R_{pot.}^{k-1}},
\end{equation} 
where \textit{Z} is the atomic number of the deformed nucleus and \textit{R$_{pot.}$} is the average radius of the potential to be deformed (i.e. the SPP). A list with the reduced transition probabilities used in the CCBA calculations is presented in Table I. \par
%
%
%
%**************Table 1***************************
\begin{table}
\begin{center}
\caption{List of the reduced transition probabilities B(E$_{k}$;0$^{+}$ $\rightarrow$ J) used in the CCBA calculations.}
\begin{tabular}{c c c }
\hline
\hline & &\\
Nucleus & \parbox{2.3cm}{\centering{B(E$_{2}$;0$^{+}$ $\rightarrow$ 2$^{+}$)}} & \parbox{2.3cm}{\centering{B(E$_{3}$;0$^{+}$ $\rightarrow$ 3$^{-}$)}}\\
 & (e$^{2}$b$^{2}$) & (e$^{2}$b$^{3}$)\\
\hline
 $^{18}$O & 0.0043$^{\dagger}$ & 0.0013$^{\star}$\\
 $^{48}$Ti & 0.072$^{\bigtriangleup}$ & 0.0074$^{\star}$\\
 $^{16}$O & - & 0.0015$^{\star}$\\
\hline
\hline
\multicolumn{3}{l}{\small ($\dagger$) from Ref. \cite{pritychenko}, ($\bigtriangleup$) from Ref. \cite{raman}, ($\star$) from Ref. \cite{kibedi}} \\
\end{tabular}
\end{center}
\end{table}
The spectroscopic amplitudes for the projectile and target overlaps were derived within the framework of nuclear shell model using the KSHELL code \cite{kshell}. For the calculation of one-neutron spectroscopic amplitudes for the $\braket{^{17}O|^{18}O}$ projectile and $\braket{^{16}O|^{17}O}$ target overlaps has been performed by using the psdmod \cite{psdmod1} effective interaction, a modified version of PSDWBT \cite{wbm}. It considers the $^{4}$He as inert core with the valence neutrons and protons in the $1p2s1d$ orbitals. This interaction gives reliable results for nuclei around $^{16}$O as testified by many of ours previous studies related to both one-proton and one-neutron transfer reaction channels \cite{spatafora_multi,linares_1n,cardozo,sgouros_titan,calabrese,ciraldo,ciraldo_1p}.\par 
As regards the spectroscopic amplitudes for the $\braket{^{28}{\rm Al}\mid^{27}{\rm Al}}$ overlaps, the the psdmod and SDPF-MU \cite{sdpfmu} interactions were used, as it was done previously for the one-proton transfer reaction on the same target \cite{sgouros_titan}, to better assess the relevance of excitations across the $1p- 2s1d$ shell-gap with respect to those across the $2s1d-1f2p$ shell-gap for the systems under investigation. In fact, at variance with psdmod, the SDPF-MU interaction is constructed in a model space spanned by the proton and neutron 2s1d and 1f2p orbitals on top of the doubly magic $^{16}$O core. It is based on the V$_{Mu}$ potential \cite{vmu_pot} and has been widely adopted to study the effects of excitations across the 2s1d-1f2p shell-gap on nuclei approaching the island of inversion at $N=20$ \cite{sdpfmu_28Mg1,sdpfmu_28Mg2,sdpfmu_28Mg3} as well as to compute the spectroscopic factors of nuclei in Ca region \cite{55ni,48ca,55sc,54ca}. As a result of our investigation (see discussion in Sec. \ref{results}), it was found that the psdmod interaction is rather inadequate to describe the magnitude of experimental data for states above $\approx$ 3.5 MeV. Then, to compute the spectroscopic amplitudes for the $\braket{^{49}{\rm Ti}|^{48}{\rm Ti}}$ overlaps the SDPF-MU interaction was employed.\par
However, due to the large model space of SDPF-MU, calculations for Al and Ti isotopes have been limited, respectively, up to 4 and 2 particle-hole excitation across the $2s1d-1f2p$ shell-gap, while all possible configurations within the $2s1d$ and $1f2p$ shells have been included. In the Appendix, a list with the calculated spectroscopic amplitudes obtained with the psdmod interaction is presented in Table II for the projectile and target overlaps, while Tables III and IV show  the spectroscopic amplitudes for the $\braket{^{28}{\rm Al}|^{27}{\rm Al}}$ and $\braket{^{49}{\rm Ti}|^{48}{\rm Ti}}$ overlaps computed with the SDFP-MU interaction.
%
%
%
%
%**************Figure 8***************************
\begin{figure}
\includegraphics[width=0.50\textwidth]{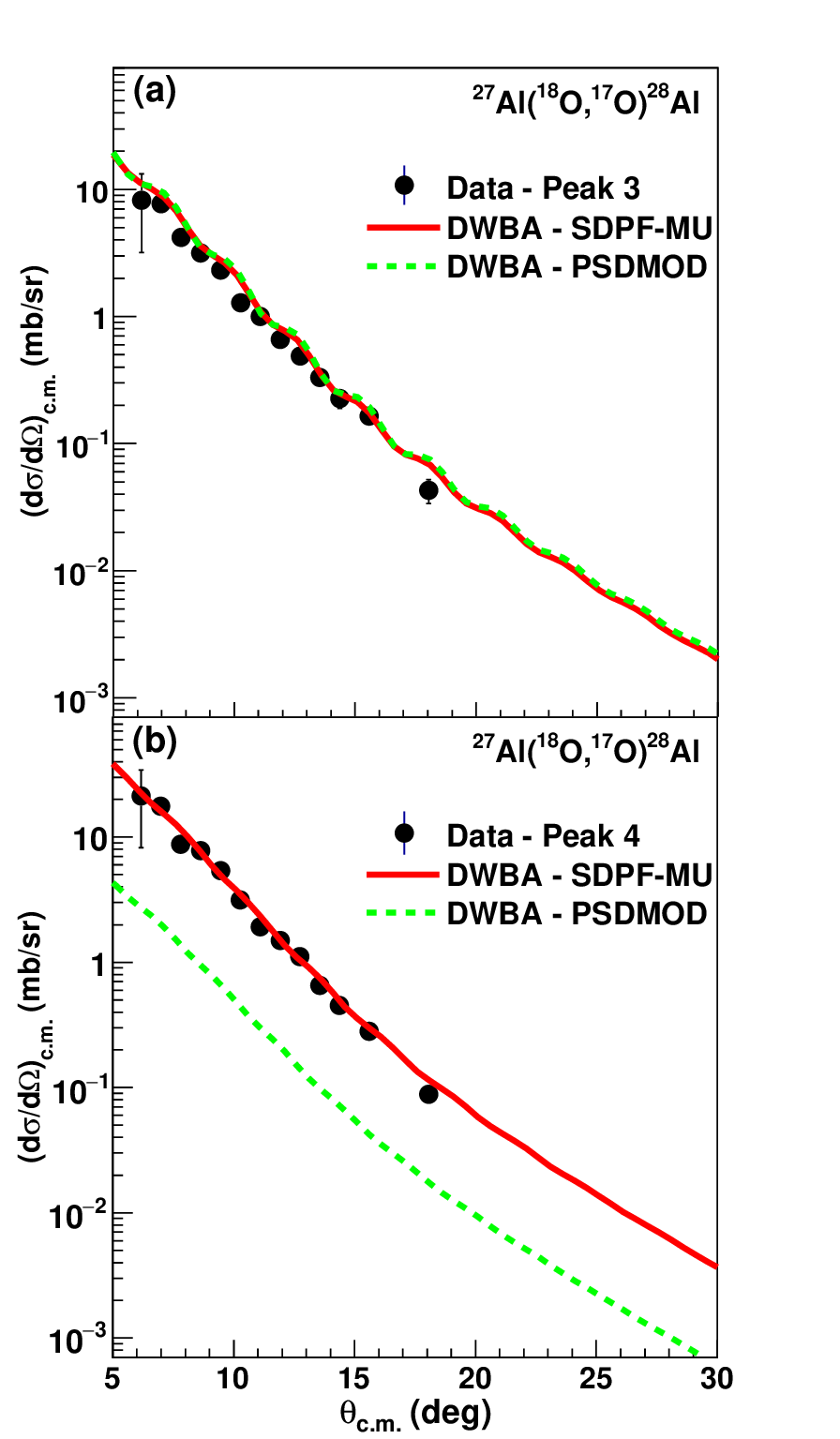}
\caption{Present angular distribution data for the $^{27}$Al($^{18}$O,$^{17}$O)$^{28}$Al reaction corresponding to (a) "Peak 3" and (b) "Peak 4" in Fig. 2b are compared to the results of DWBA calculations which were performed adopting two different sets of spectroscopic amplitudes.}
\end{figure}
%
%
%
%
%
%**************Figure 9***************************
\begin{figure}
\includegraphics[width=0.50\textwidth]{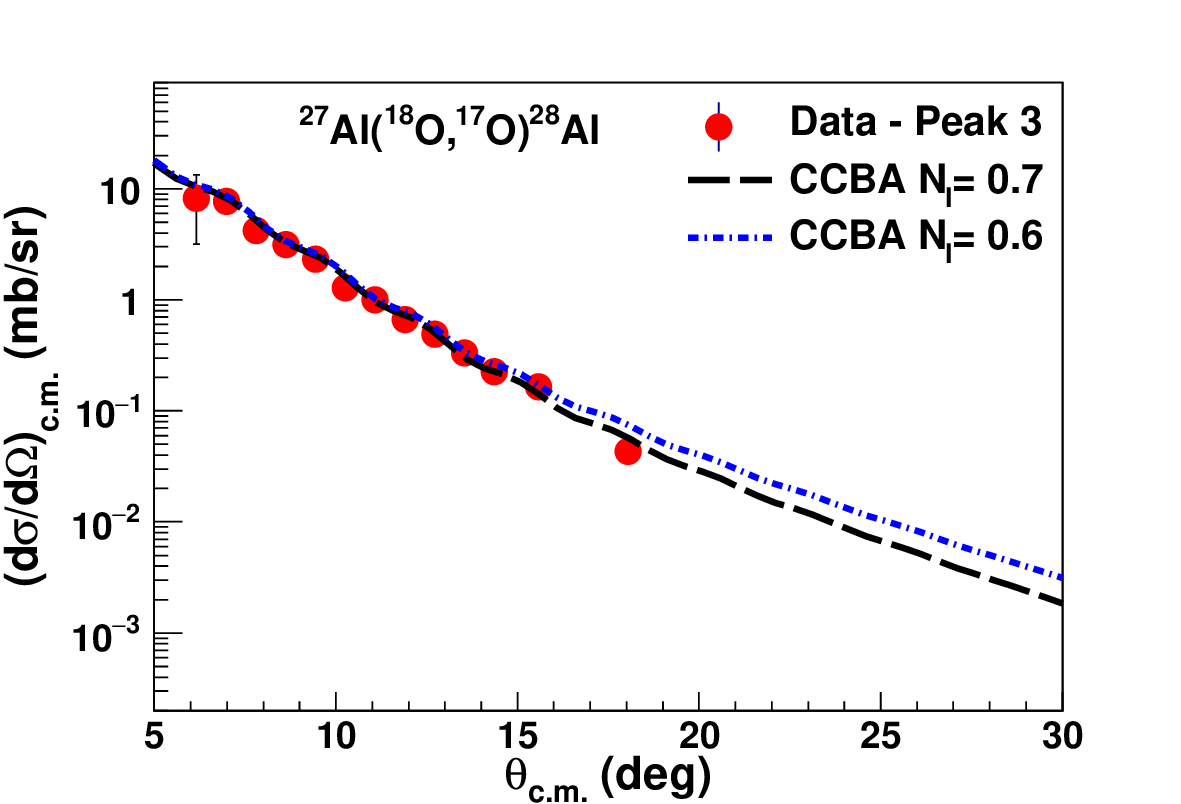}
\caption{Present angular distribution data for the $^{27}$Al($^{18}$O,$^{17}$O)$^{28}$Al reaction corresponding to "Peak 3" in Fig.2b are compared to the results of CCBA calculations which were performed adopting two different normalization coefficients for the imaginary part of the optical potential at the entrance channel.}
\end{figure}
\section{Results and discussion}
\label{results}
Starting with the case of the $^{16}$O($^{18}$O,$^{17}$O)$^{17}$O reaction, as it was already mentioned the reaction products at the exit channel are identical nuclei. This fact may give rise to interference effects owing to the two-body nature of the reaction, that is the ejectiles at forward angles cannot be distinguished kinematically from the recoil nuclei. Therefore, the theoretical cross-sections were obtained from the coherent sum of the scattering amplitudes f($\theta$) and f(\ang{180}-$\theta$). The resulted theoretical curves are compared to the experimental data in Fig.5, showing a fair agreement. The CCBA calculations indicate a weak coupling to collective states of the projectile and target, with a moderate coupling influence to be inferred for the cross-sections corresponding to "Peak 2". The errors in the differential cross-section data, which include the background subtraction uncertainty, range from (7 to 21)\%, (14 to 31)\% and (12 to 34)\% for the data of "Peak 1", "Peak 2" and "Peak 3", respectively. These values are dominated by the uncertainties in the background subtraction, although a residual error of about 50\% of the initial one remains, when this source of uncertainty is switched-off from the error propagation. Overall, the data interpretation is rather satisfactory. In more details, the data in Peaks 1 and 2 are slightly overestimated from theory by a factor of $\approx$ 1.2, while the experimental cross-sections in the region of "Peak 3" are underestimated by a factor of $\approx$ 0.7.\par
As mentioned above, for the $^{27}$Al($^{18}$O,$^{17}$O)$^{28}$Al reaction, we have computed the theoretical cross-sections using both the psdmod and SDPF-MU interactions. It was found that at low excitations energies, namely for the first three peaks in Fig.2b, the theoretical cross-sections computed with the two effective interactions are equivalent and describe well the experimental results, as it can be seen in Fig.8a for instance for "Peak 3". This is not surprising. In fact, correlations  within the $2s1d$ shell are dominant at low-energy and the matrix elements involving the orbitals of this shell are based on the USD Hamiltonian \cite{USD} for both interactions.\par
On the other hand, the two interactions provide different results for "Peak 4" and "Peak 5". In these two energy regions, psdmod calculations significantly underestimate the experimental values, while a good agreement is obtained with the SDPF-MU interaction. This is illustrated for "Peak 4" in Fig.8b, where theoretical angular distributions in DWBA obtained with both interactions are compared to the experimental data. The reason of this difference can be traced back to the important effects of sd-fp cross-shell excitations in the energy region above $\approx$ 3.5 MeV. As a matter of fact, the main contributions to the cross-sections corresponding to "Peak 4" and "Peak 5" arise from negative-parity states which can be constructed only by including cross-shell excitations. The better agreement obtained with SDPF-MU indicates that a more crucial role is played by the $2p-1f$ orbitals. This was also highlighted in the $^{27}$Al(d,p)$^{28}$Al transfer studies in Refs. \cite{alum28_carola,28al_freeman}, which underline in particular the role of the $(1d^{-1}_{5/2}1f)$ configuration in contributing to the high-lying negative-parity states.\par
The result of this analysis highlights the selectivity of heavy-ion induced transfer reactions on the details of the adopted model space confirming the results of the work by Ciraldo et al. \cite{ciraldo}, claiming for a sizeable sensitivity of such kind of transfer reactions on the adopted nuclear structure model. The description of $^{28}$Al nucleus favors the use of the SDPF-MU interaction, a fact that should be taken into account in case where strong single-particle states of this nucleus can serve as intermediate steps for second order processes.\par
%
%though significant, is small
%
Having established the validity of the SDPF-MU interaction, a comparison between the complete set of experimental data and theoretical calculations is presented in Fig.3. From an inspection on the angular distributions, it is evident that the DWBA calculations describe very well the shape and the magnitude of the experimental data, signaling that the populated final states are characterized by a substantial single-particle strength. This is well borne out by the results of the CCBA calculations, where the inclusion of couplings to the 2$_{1}^{+}$ and 3$_{1}^{-}$ collective states of projectile has a small impact on the predicted cross-sections. Couplings to exited states of $^{27}$Al target were not considered in the calculations and thus, the imaginary part of the optical potential at the entrance partition was scaled by a factor of 0.70 instead of 0.60 which is the standard prescription for the SPP \cite{spp3}. To further elucidate the effect on the choice of N$_{I}$, exploratory calculations adopting the standard value of the normalization coefficient were performed and the results are visualized in Fig.9. Using as benchmark the data of Peak 3, it was found that the angular distributions cross-sections in the measured angular range are rather insensitive to the choice of N$_{I}$, while a small change in the slope of the theoretical curves is observed with the increasing angle. Similar conclusions were also drawn from the analysis of the rest of the available data. Overall, a very good agreement is observed between theoretical predictions and experimental data. The calculations slightly overestimate the data by a factor ranging from 1.0 to 1.15, but the case of Peak 3 where a factor of 1.4 is found.\par
%
%
%
%
%**************Figure 10***************************
\begin{figure}
\includegraphics[width=0.50\textwidth]{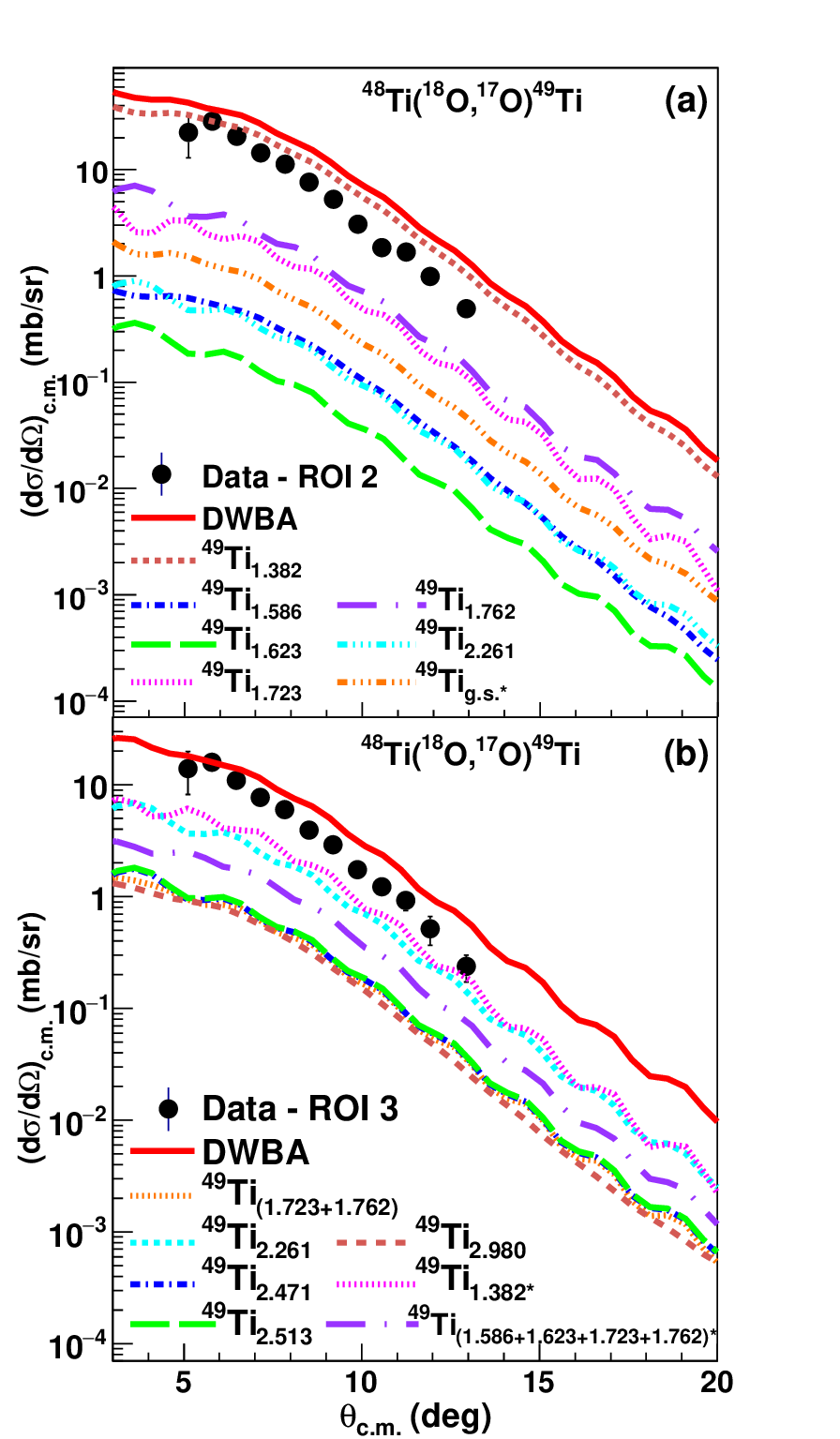}
\caption{Present angular distribution data for the $^{48}$Ti($^{18}$O,$^{17}$O)$^{49}$Ti reaction corresponding to (a) "ROI 2" and (b) "ROI 3" in Fig. 6b are compared to the results of DWBA calculations. Theoretical angular distribution cross-sections for the transitions to the involved states of the ejectile and the residual nuclei are presented with the colored curves. In the legend, each curve is labeled by the corresponding excitation energy of the recoil nucleus for transitions to the g.s. of the ejectile and an asterisk in case where $^{17}$O ejectile is excited to the \(\frac{1}{2}\)$_{1}^{+}$ (0.871 MeV) state. The sum of all transitions is illustrated by the red solid line. For ROI 3, only the 7 most important contributions are shown.}
\end{figure}
Let us now discuss the $^{48}$Ti($^{18}$O,$^{17}$O)$^{49}$Ti reaction. The experimental data are presented in Fig. 7. Since the majority of the extracted yields is a result of an $^{27}$Al background subtraction, the errors in the data points are dominated from the background subtraction uncertainties. In more details, the errors in the differential cross-section data, which includes the background subtraction uncertainty, range from (5 to 10)\%, (8 to 30)\% and (9 to 20)\% for the data of ROI 2, ROI 3 and ROI 4, respectively. In this case, a residual error of about (30-40)\% of the initial error value remains, if the background subtraction uncertainty is switched-off from the error propagation. The errors for the data of ROI 1, which is free of any background, are less than 10\%.\par
In general, for all the reactions under study the overall errors are typically smaller than the observed variations in the predicted cross-sections from different state-of-art theoretical models. Thus, we may claim to have enough sensitivity to allow a valuable discrimination of different nuclear structure and reaction models.\par
The experimental data are compared to theoretical prediction in Fig. 7. Due to the large number of states introduced in the calculations the total theoretical cross-sections are reported for each energy region, but the individual contribution for each transition is given explicitly in Fig. 10 only for ROI 2 and ROI 3, where the calculated cross-sections overestimate the measured ones by a factor of $\approx$ 1.6 (see Table V). To a leaser extent, the same observation holds also for ROI 1 where the relevant factor is reduced to $\approx$ 1.2. As regards the ROI 4 the experimental data are in good agreement with the results of the DWBA and CCBA calculations (a factor of $\approx$ 0.9 is found).\par
The discrepancies met in ROI 2 and ROI 3 could be explained to some extent by considering the truncation scheme adopted in the calculations of Ti isotopes, which includes at most 2 particle-hole configurations. However, to further elucidate this point it may be useful to analyze Fig. 10.\par 
As a matter of fact, in ROI 2 the discrepancy can be attributed to the large cross-section predicted for the transition to the \(\frac{3}{2}\)$_{1}^{-}$ state at 1.382 MeV, whose contribution alone already overestimates the magnitude of the experimental data. However, the calculated spectroscopic amplitude for this state is consistent with the value suggested from $(d,p)$ experiments \cite{49ti_dp_reaction}, which may indicate that the single-particle component of this state is overestimated. Furthermore, we should bear in mind that the spectroscopic factors obtained in \cite{49ti_dp_reaction} were determined by normalizing the DWBA predictions to the experimental data, so are in principle susceptible to the experimental uncertainties as well as to the details of the reaction model (e.g. optical potentials).\par
As regards ROI 3, it is observed that the two main contributions arise from the transitions to the \(\frac{3}{2}\)$_{1}^{-}$ state as in ROI 2 and to the \(\frac{5}{2}\)$_{3}^{-}$ state at 2.261 MeV. From the previous discussion, it is expected that the contribution from the \(\frac{3}{2}\)$_{1}^{-}$ state is rather overestimated. Moreover, for the 2.261 MeV state a higher spectroscopic amplitude is predicted by our calculations as compared to the experimental value (0.36 versus 0.14).\par
We may conclude that a slightly different distribution of the single-neutron strength is needed for the \(\frac{3}{2}\)$_{1}^{-}$ and \(\frac{5}{2}\)$_{3}^{-}$ states than that provided by the SDFP-MU interaction. This seems to point to some deficiencies in the matrix elements of this interaction concerning configurations of $1f2p$ shell. In fact, $2s1d$ and cross-shell matrix elements, whose reliability has been even testified by calculations for Al isotopes, do not play a relevant role as evidenced by comparing results with and without cross-shell excitations that are found to be very similar. Furthermore, by using the KB3 effective interaction \cite{kb3}, defined only in the $1f2p$ space, smaller spectroscopic amplitudes are predicted for the \(\frac{3}{2}\)$_{1}^{-}$ and \(\frac{5}{2}\)$_{3}^{-}$ states and the theoretical angular distributions cross-sections for the ROI 2 and ROI 3 provide a better description of the experimental data, especially for ROI 3 (see Fig. 7). A tentative calculation using a value of $\approx$ 0.40 for the spectroscopic amplitude for the \(\frac{3}{2}\)$_{1}^{-}$ state was performed. It was found that using such a value the angular distribution data of both ROIs 2 and 3 can be described reasonably by theory. However, this is an arbitrary choice and therefore, the obtained results are not included in the present work.\par
The above discussion indicated that the SDPF-MU interaction would require adjustments of some matrix elements. However, given the success of this interaction to explain the structure of $^{47}$Sc nucleus including the one-proton transfer cross-sections on $^{48}$Ti \cite{sgouros_titan} and in prospect of a systematic study, we may conclude that the SDPF-MU interaction with the present truncation scheme provides a reasonable description of the data on the $^{48}$Ti($^{18}$O,$^{17}$O)$^{49}$Ti reaction.\par
\section{Summary and conclusions}
Absolute differential cross-sections measurements for the ($^{18}$O,$^{17}$O) one-neutron transfer reaction on three different targets namely, $^{27}$Al, $^{16}$O and $^{48}$Ti were performed at the energy of 275 MeV. Angular distributions of the $^{17}$O$^{8+}$ reaction ejectiles were determined in a wide angular range by means of the MANGEX large acceptance magnetic spectrometer at INFN-LNS. The differential cross-section data were analyzed under the DWBA framework and overall a good agreement between experimental and theoretical cross-sections is inferred. Moreover, CCBA calculations were also performed suggesting a weak coupling to collective states of the projectile and target nuclei.\par
Further on, the sensitivity of the calculated cross-sections on the configuration space used in the large-scale shell model calculation was sought in the data obtained with the $^{27}$Al and $^{48}$Ti targets. For the $^{27}$Al($^{18}$O,$^{17}$O)$^{28}$Al reaction, it was found that the inclusion of $fp$ shell in the adopted model space is mandatory for the description of the negative parity states of $^{28}$Al (see Fig.8). Moreover, for the case of the $^{48}$Ti($^{18}$O,$^{17}$O)$^{49}$Ti reaction it was found that the discrepancies met between experimental data and theoretical predictions (see Fig.7) can be ascribed, to some extent, to the adopted truncation of the shell model basis. The most striking example was the case of ROI 3, where a clear improvement between data and theory is observed when a full fp shell model basis is employed instead of a truncated one. However, for the rest of the data the differences when using a truncated or a full fp shell model basis are less evident.\par
Finally, the results of the present work are complementary to those reported for the one-proton transfer reaction for the same system. Both results will help to clarify the degree of competition between sequential nucleon transfer and the direct meson exchange dynamics in single and double charge exchange reactions, which is a relevant aspect under scrutiny by nuclear reaction theory. Moreover, both one-neutron and one-proton transfer data sets can be used as guidelines to constrain the reaction models for the proper description of two-neutron and two-proton transfer reaction mechanisms, respectively, which constitute a piece of the puzzle for the description of the complete DCE mechanism.
\section*{Acknowledgments}
The authors would like to thank the staff of the LNS Accelerator for the support during the experiment. The research leading to these results was partially funded by the European Research Council (ERC) under the European Union’s Horizon 2020 Research and Innovation Programme (Grant Agreement No 714625). We also acknowledge the CINECA award under the ISCRA initiative (code HP10B51E4M) and through the INFN-CINECA agreement for the availability of high performance computing resources and support. One of us (G.D.G.) acknowledges the support by the funding program VALERE of Univerit\`a degli Studi della Campania "Luigi Vanvitelli". We also acknowledge partial financial support from CNPq, FAPERJ, FAPESP (proc. no. 2019/07767-1) and Instituto Nacional de Ci\^{e}ncia e Tecnologia - F\'isica Nuclear e Aplica\c{c}\~{o}es (INCT-FNA, proc. no. 464898/2014-5), Brazil.
\section{Appendix}
%
%
%
%**************Table 2***************************
\begin{table}[ht]
\begin{center}
\caption{One-neutron spectroscopic amplitudes for the projectile and target overlaps, calculated using the psdmod interaction. The symbols \textbf{n}, \textbf{l} and \textbf{j} correspond to the principal quantum number, the orbital and the total angular momentum of the transferred neutron orbitals, respectively.}
\begin{tabular}{|c| c|c| c |}
\hline
Initial state & \parbox{1.3cm}{\centering{\textbf{nlj}}} & \parbox{2.5cm}{\centering{Final state}} & \parbox{2.0cm}{\centering{Spectroscopic amplitude}}\\
\hline
 &	      &				    &	     \\
 & 1d$_{5/2}$ & $^{17}$O$_{g.s.}$ (5/2$^+$) & 1.2708\\
 $^{18}$O$_{g.s.}$ (0$^+$) & 2s$_{1/2}$ & $^{17}$O$_{0.871}$ (1/2$^+$) & -0.4345\\
 & 1p$_{1/2}$ & $^{17}$O$_{3.055}$ (1/2$^-$) & 0.8155\\
 &	      &				    &	     \\
 & 1d$_{5/2}$ & $^{17}$O$_{g.s.}$ (5/2$^+$) & -1.0734\\
 & 1d$_{3/2}$ & $^{17}$O$_{g.s.}$ (5/2$^+$) & -0.0799\\
 & 2s$_{1/2}$ & $^{17}$O$_{g.s.}$ (5/2$^+$) & -0.5093\\
 & 1d$_{5/2}$ & $^{17}$O$_{0.871}$ (1/2$^+$) & 0.4994\\
$^{18}$O$_{1.982}$ (2$^+$) & 1d$_{3/2}$ & $^{17}$O$_{0.871}$ (1/2$^+$) & 0.1737\\
 & 1p$_{3/2}$ & $^{17}$O$_{3.055}$ (1/2$^-$) & -0.0207\\
 & 1p$_{3/2}$ & $^{17}$O$_{3.843}$ (5/2$^-$) & -0.0085\\
 & 1p$_{1/2}$ & $^{17}$O$_{3.843}$ (5/2$^-$) & -0.6787\\
 &	      &				    &	     \\
 & 1p$_{3/2}$ & $^{17}$O$_{g.s.}$ (5/2$^+$) &  0.0988\\
 & 1p$_{1/2}$ & $^{17}$O$_{g.s.}$ (5/2$^+$) & -0.1135\\
 & 1d$_{5/2}$ & $^{17}$O$_{3.055}$ (1/2$^-$) & -0.2430\\
 $^{18}$O$_{5.098}$ (3$^-$) & 1d$_{5/2}$ & $^{17}$O$_{3.843}$ (5/2$^-$) & 0.5860\\
 & 1d$_{3/2}$ & $^{17}$O$_{3.843}$ (5/2$^-$) & 0.0888\\
 & 2s$_{1/2}$ & $^{17}$O$_{3.843}$ (5/2$^-$) & 0.1220\\
 &	      &				    &	     \\
 & 1d$_{5/2}$ & $^{17}$O$_{g.s.}$ (5/2$^+$) & 0.9445\\
 $^{16}$O$_{g.s.}$ (0$^+$) & 2s$_{1/2}$ & $^{17}$O$_{0.871}$ (1/2$^+$) & -0.9633\\
 & 1p$_{1/2}$ & $^{17}$O$_{3.055}$ (1/2$^-$) & -0.2810\\
 &	      &				    &	     \\
 & 1p$_{3/2}$ & $^{17}$O$_{g.s.}$ (5/2$^+$) & -0.1974\\
 & 1p$_{1/2}$ & $^{17}$O$_{g.s.}$ (5/2$^+$) & 0.5993\\
 & 1d$_{5/2}$ & $^{17}$O$_{3.055}$ (1/2$^-$) & -0.7256\\
 $^{16}$O$_{6.130}$ (3$^-$) & 1d$_{5/2}$ & $^{17}$O$_{3.843}$ (5/2$^-$) & -0.9309\\
 & 1d$_{3/2}$ & $^{17}$O$_{3.843}$ (5/2$^-$) & 0.0768\\
 & 2s$_{1/2}$ & $^{17}$O$_{3.843}$ (5/2$^-$) & -0.2755\\
\hline
\end{tabular}
\end{center}
\end{table}
%
%
%
%**************Table 3***************************
\LTcapwidth=\linewidth
\begin{longtable}{|c |c| c |c| }
\caption{One-neutron spectroscopic amplitudes for the $\braket{^{28}Al|^{27}Al}$ overlaps calculated using the SDPF-MU interaction. The symbols \textbf{n}, \textbf{l} and \textbf{j} correspond to the principal quantum number, the orbital and the total angular momentum of the transferred neutron orbitals, respectively.}\label{tab:spectr_ampl}\\
            \hline
Initial state & \parbox{1.1cm}{\centering{\textbf{nlj}}} & \parbox{2.2cm}{\centering{Final state}} & \parbox{2.0cm}{\centering{Spectroscopic amplitude}}\\
            \hline
            \endfirsthead
            %
            %intestazione normale
            \multicolumn{4}{l}{\makebox[2.5cm][l]{TABLE III: continued}} \\
      %      \toprule[1pt]\midrule[0.3pt]
            \hline
Initial state & \parbox{1.1cm}{\centering{\textbf{nlj}}} & \parbox{2.2cm}{\centering{Final state}} & \parbox{2.0cm}{\centering{Spectroscopic amplitude}}\\
            \hline
            \endhead
\multirow{11}{*}{$^{27}$Al$_{g.s.}$ (5/2$^+$)} & 1d$_{5/2}$  & \multirow{3}{*}{ $^{28}$Al$_{g.s.}$ (3$^+$)} &  0.3247\\
& 1d$_{3/2}$ &                             & -0.2148\\
& 2s$_{1/2}$ &                             &  0.5487\\
&	         &				               &	     \\
& 1d$_{5/2}$ & \multirow{3}{*}{ $^{28}$Al$_{0.031}$ (2$^+$)} & -0.2607\\
& 1d$_{3/2}$ &                             &  0.4280\\
& 2s$_{1/2}$ &                              & -0.3471\\
&	         &				               &	     \\
& 1d$_{5/2}$ & $^{28}$Al$_{0.972}$ (0$^+$) & -0.5548\\
&	         &				               &	     \\
& 1d$_{5/2}$ & \multirow{1}{*}{ $^{28}$Al$_{1.014}$ (3$^+$)} & -0.0090\\\cline{1-4}
& 1d$_{3/2}$  & \multirow{2}{*}{ $^{28}$Al$_{1.014}$ (3$^+$)} &  0.4055\\
& 2s$_{1/2}$ &                             &  0.2719\\
&	         &				               &	     \\
\multirow{58}{*}{$^{27}$Al$_{g.s.}$ (5/2$^+$)} & 1d$_{5/2}$ & \multirow{2}{*}{ $^{28}$Al$_{1.373}$ (1$^+$)} & -0.2881\\
& 1d$_{3/2}$ &                              &  0.5582\\
&	         &				               &	     \\
& 1d$_{5/2}$ & \multirow{2}{*}{ $^{28}$Al$_{1.620}$ (1$^+$)} &  0.3908\\
& 1d$_{3/2}$ &                             &  0.1032\\
&	         &				               &	     \\
& 1d$_{5/2}$ & \multirow{3}{*}{ $^{28}$Al$_{1.623}$ (2$^+$)} & -0.4035\\
& 1d$_{3/2}$ &                             & -0.2405\\
& 2s$_{1/2}$ &                              & -0.0538\\
&	         &				               &	     \\
& 1d$_{5/2}$ & \multirow{3}{*}{ $^{28}$Al$_{2.139}$ (2$^+$)} &  0.0900\\
& 1d$_{3/2}$ &                             &  0.0391\\
& 2s$_{1/2}$ &                             & -0.2229\\
&	         &				               &	     \\
& 1d$_{5/2}$ & \multirow{2}{*}{ $^{28}$Al$_{2.201}$ (1$^+$)} & -0.0811\\
& 1d$_{3/2}$ &                             & -0.0320\\
&	         &				               &	     \\
& 1d$_{5/2}$ & \multirow{2}{*}{ $^{28}$Al$_{2.272}$ (4$^+$)} & -0.4929\\
& 1d$_{3/2}$ &                             & -0.0285\\
&	         &				               &	     \\
& 1d$_{5/2}$ & \multirow{3}{*}{ $^{28}$Al$_{2.486}$ (2$^+$)} &  0.1128\\
& 1d$_{3/2}$ &                             & -0.4223\\
& 2s$_{1/2}$ &                             & -0.2596\\
&	         &				               &	     \\
& 1d$_{5/2}$ & $^{28}$Al$_{2.582}$ (5$^+$) & -0.4314\\
&	         &				               &	     \\
& 1d$_{5/2}$ & \multirow{2}{*}{ $^{28}$Al$_{2.656}$ (4$^+$)} & -0.1635\\
& 1d$_{3/2}$ &                             & -0.5499\\
&	         &				               &	     \\
%\multicolumn{4}{l}{} \\
& 1d$_{5/2}$ & \multirow{3}{*}{ $^{28}$Al$_{3.347}$ (2$^+$)}  & 0.0500\\
& 1d$_{3/2}$ &                             & -0.0248\\
& 2s$_{1/2}$ &                             & -0.1883\\
&	         &				               &	     \\
& 1f$_{7/2}$ & \multirow{3}{*}{ $^{28}$Al$_{3.465}$ (4$^-$)} & -0.6114\\
& 1f$_{5/2}$ &                             &  0.0224\\
& 2p$_{3/2}$ &                             &  0.3764\\
&	         &				               &	     \\
& 1f$_{7/2}$ & \multirow{4}{*}{ $^{28}$Al$_{3.591}$ (3$^-$)} & -0.3513\\
& 1f$_{5/2}$ &                             & -0.0064\\
& 2p$_{3/2}$ &                             &  0.3741\\
& 2p$_{1/2}$ &                             & -0.0579\\
&	         &				               &	     \\
& 1d$_{5/2}$ & $^{28}$Al$_{3.760}$ (0$^+$) & -0.1401\\
&	         &				               &	     \\
& 1f$_{7/2}$ & \multirow{4}{*}{ $^{28}$Al$_{3.876}$ (2$^-$)} & -0.3975\\
& 1f$_{5/2}$ &                             & -0.0399\\
& 2p$_{3/2}$ &                             &  0.2946\\
& 2p$_{1/2}$ &                             & -0.0503\\
&	         &				               &	     \\
& 1f$_{7/2}$ & \multirow{2}{*}{ $^{28}$Al$_{4.033}$ (5$^-$)} &  0.7382\\
& 1f$_{5/2}$ &                              & -0.0799\\
&	         &				               &	     \\
& 1f$_{7/2}$ & \multirow{4}{*}{$^{28}$Al$_{4.691}$ (3$^-$)} & -0.3297 \\
& 1f$_{5/2}$ &                              & -0.0267\\
& 2p$_{3/2}$ &                              &  0.4658\\
& 2p$_{1/2}$ &                              & -0.0709\\
&	         &				                &	     \\\cline{1-4}
\multicolumn{4}{l}{} \\
\multirow{21}{*}{$^{27}$Al$_{g.s.}$ (5/2$^+$)} & 1f$_{7/2}$ & \multirow{4}{*}{$^{28}$Al$_{4.765}$ (2$^-$)} & -0.2452\\
& 1f$_{5/2}$ &                              & -0.0323\\
& 2p$_{3/2}$ &                              &  0.4940\\
& 2p$_{1/2}$ &                              &  0.1309\\
&	         &				               &	     \\
%\multicolumn{4}{l}{} \\
%\multicolumn{4}{l}{} \\
& 1f$_{7/2}$ & \multirow{4}{*}{$^{28}$Al$_{4.904}$ (2$^-$)} &  0.1721\\
& 1f$_{5/2}$ &                             & -0.0249\\
& 2p$_{3/2}$ &                             &  0.2951\\
& 2p$_{1/2}$ &                             &  0.0408\\
&	         &				               &	     \\
& 1f$_{7/2}$ & \multirow{4}{*}{$^{28}$Al$_{4.997}$ (2$^-$)} &  0.1396\\
& 1f$_{5/2}$ &                             & -0.0613\\
& 2p$_{3/2}$ &                             &  0.3182\\
& 2p$_{1/2}$ &                             &  0.3931\\
&	         &				               &	     \\
& 1f$_{7/2}$ & \multirow{4}{*}{$^{28}$Al$_{5.135}$ (3$^-$)} & -0.2723\\
& 1f$_{5/2}$ &                              &  0.0164\\
& 2p$_{3/2}$ &                              & -0.4613\\
& 2p$_{1/2}$ &                              &  0.1493\\
&	         &				               &	     \\
& 1f$_{7/2}$ & \multirow{1}{*}{$^{28}$Al$_{5.165}$ (6$^-$)} &  0.6523\\
\hline
\end{longtable}
%
%
%
%
%**************Table 4***************************
\LTcapwidth=\linewidth
\begin{longtable}{|c| c| c |c|}
\caption{One-neutron spectroscopic amplitudes for the $\braket{^{49}Ti|^{48}Ti}$ overlaps calculated using the SDPF-MU interaction. The symbols \textbf{n}, \textbf{l} and \textbf{j} correspond to the principal quantum number, the orbital and the total angular momentum of the transferred neutron orbitals, respectively.}\label{tab:spectr_ampl2}\\
            \hline
Initial state & \parbox{1.1cm}{\centering{\textbf{nlj}}} & \parbox{2.2cm}{\centering{Final state}} & \parbox{2.0cm}{\centering{Spectroscopic amplitude}}\\
            \hline
            \endfirsthead
            %
            %intestazione normale
            \multicolumn{4}{l}{\makebox[2.5cm][l]{TABLE IV: continued}} \\
      %      \toprule[1pt]\midrule[0.3pt]
            \hline
Initial state & \parbox{1.1cm}{\centering{\textbf{nlj}}} & \parbox{2.2cm}{\centering{Final state}} & \parbox{2.0cm}{\centering{Spectroscopic amplitude}}\\
            \hline
            \endhead
\multirow{17}{*}{$^{48}$Ti$_{g.s.}$ (0$^+$)} & 1f$_{7/2}$  & $^{49}$Ti$_{g.s.}$ (7/2$^-$) &  0.4872\\
& 2p$_{3/2}$ &  $^{49}$Ti$_{1.382}$ (3/2$^-$) & -0.8217\\
& 2p$_{3/2}$ &  $^{49}$Ti$_{1.586}$ (3/2$^-$) & -0.1145\\
& 1f$_{5/2}$ &  $^{49}$Ti$_{1.623}$ (5/2$^-$) &  0.0778\\
& 2p$_{1/2}$ &  $^{49}$Ti$_{1.723}$ (1/2$^-$) & -0.4762\\
& 1f$_{5/2}$ &  $^{49}$Ti$_{1.762}$ (5/2$^-$) &  0.3684\\
& 1f$_{5/2}$ &  $^{49}$Ti$_{2.261}$ (5/2$^-$) &  0.3570\\
& 1f$_{5/2}$ &  $^{49}$Ti$_{2.471}$ (5/2$^-$) & -0.1689\\
& 1f$_{5/2}$ & $^{49}$Ti$_{2.513}$ (5/2$^-$) &  0.1710\\
& 1f$_{7/2}$ &  $^{49}$Ti$_{2.980}$ (7/2$^-$) &  0.1607\\
& 1f$_{7/2}$ &  $^{49}$Ti$_{3.042}$ (7/2$^-$) &  0.1348\\
& 2p$_{1/2}$ &  $^{49}$Ti$_{3.175}$ (1/2$^-$) & -0.3715\\
& 2p$_{3/2}$ &  $^{49}$Ti$_{3.261}$ (3/2$^-$) & -0.2505\\
& 2p$_{3/2}$ &  $^{49}$Ti$_{3.428}$ (3/2$^-$) & -0.0151\\
& 2p$_{1/2}$ &  $^{49}$Ti$_{3.469}$ (1/2$^-$) & -0.4048\\
& 1f$_{5/2}$ &  $^{49}$Ti$_{3.511}$ (5/2$^-$) &  0.0823\\
& 1f$_{5/2}$ &  $^{49}$Ti$_{3.618}$ (5/2$^-$) & -0.0737\\
& 2p$_{3/2}$ &  $^{49}$Ti$_{3.788}$ (3/2$^-$) & -0.0416\\
& 1f$_{5/2}$ &  $^{49}$Ti$_{3.855}$ (5/2$^-$) & -0.1019\\\cline{1-4}
\multirow{10}{*}{$^{48}$Ti$_{0.984}$ (2$^+$)} & 1f$_{7/2}$ & \multirow{3}{*}{ $^{49}$Ti$_{g.s.}$ (7/2$^-$)} &  0.8999\\
& 1f$_{5/2}$ &                             & -0.0423\\
& 2p$_{3/2}$ &                             & -0.0796\\
&	         &				               &	     \\
& 1f$_{7/2}$ & \multirow{4}{*}{ $^{49}$Ti$_{1.382}$ (3/2$^-$)} &  -0.2767\\
& 1f$_{5/2}$ &                             & -0.0171\\
& 2p$_{3/2}$ &                             &  0.0069\\
& 2p$_{1/2}$ &                             & -0.1856\\
&	         &				               &	     \\
& 1f$_{7/2}$ & \multirow{2}{*}{ $^{49}$Ti$_{1.586}$ (3/2$^-$)} &  -0.2823\\
& 1f$_{5/2}$ &                             & -0.0913\\\cline{1-4}
\multirow{60}{*}{$^{48}$Ti$_{0.984}$ (2$^+$)} & 2p$_{3/2}$ & \multirow{2}{*}{ $^{49}$Ti$_{1.586}$ (3/2$^-$)} &  0.6818\\
& 2p$_{1/2}$ &                             & -0.1317\\
&	         &				               &	     \\
& 1f$_{7/2}$ & \multirow{4}{*}{ $^{49}$Ti$_{1.623}$ (5/2$^-$)} &   0.1199\\
& 1f$_{5/2}$ &                             & -0.1316\\
& 2p$_{3/2}$ &                             & -0.4205\\
& 2p$_{1/2}$ &                             &  0.0823\\
&	         &				               &	     \\
& 1f$_{5/2}$ & \multirow{2}{*}{ $^{49}$Ti$_{1.723}$ (1/2$^-$)} &   0.2164\\
& 2p$_{3/2}$ &                             &  0.6434\\
&	         &				               &	     \\
& 1f$_{7/2}$ & \multirow{4}{*}{ $^{49}$Ti$_{1.762}$ (5/2$^-$)} &   0.0894\\
& 1f$_{5/2}$ &                             & -0.2747\\
& 2p$_{3/2}$ &                             & -0.4277\\
& 2p$_{1/2}$ &                             & -0.3669\\
&	         &				               &	     \\
& 1f$_{7/2}$ & \multirow{4}{*}{ $^{49}$Ti$_{2.261}$ (5/2$^-$)} &  -0.0089\\
& 1f$_{5/2}$ &                             & -0.3213\\
& 2p$_{3/2}$ &                             &  0.5434\\
& 2p$_{1/2}$ &                             &  0.0465\\
&	         &				               &	     \\
& 1f$_{7/2}$ & \multirow{4}{*}{ $^{49}$Ti$_{2.471}$ (5/2$^-$)} &  -0.2189\\
& 1f$_{5/2}$ &                             &  0.1602\\
& 2p$_{3/2}$ &                             & -0.1679\\
& 2p$_{1/2}$ &                             &  0.1079\\
&	         &				               &	     \\
& 1f$_{7/2}$ & \multirow{4}{*}{ $^{49}$Ti$_{2.513}$ (5/2$^-$)} &   0.0162\\
& 1f$_{5/2}$ &                             & -0.1138\\
& 2p$_{3/2}$ &                             & -0.2825\\
& 2p$_{1/2}$ &                             &  0.4077\\
&	         &				               &	     \\
& 1f$_{7/2}$ & \multirow{3}{*}{ $^{49}$Ti$_{2.980}$ (7/2$^-$)} &  -0.0379\\
& 1f$_{5/2}$ &                             &  0.0193\\
& 2p$_{3/2}$ &                             &  0.6600\\
&	         &				               &	     \\
& 1f$_{7/2}$ & \multirow{3}{*}{ $^{49}$Ti$_{3.042}$ (7/2$^-$)} &  -0.2316\\
& 1f$_{5/2}$ &                             & -0.0843\\
& 2p$_{3/2}$ &                             &  0.4345\\
&	         &				               &	     \\
& 1f$_{5/2}$ & \multirow{2}{*}{ $^{49}$Ti$_{3.175}$ (1/2$^-$)} &  -0.3173\\
& 2p$_{3/2}$ &                             & -0.0482\\
&	         &				               &	     \\
& 1f$_{7/2}$ & \multirow{4}{*}{ $^{49}$Ti$_{3.261}$ (3/2$^-$)} &  -0.1361\\
& 1f$_{5/2}$ &                             & -0.1135\\
& 2p$_{3/2}$ &                             & -0.4522\\
& 2p$_{1/2}$ &                             &  0.0815\\
&	         &				               &	     \\
& 1f$_{7/2}$ & \multirow{4}{*}{ $^{49}$Ti$_{3.428}$ (3/2$^-$)} &  -0.1108\\
& 1f$_{5/2}$ &                             & -0.1655\\
& 2p$_{3/2}$ &                             & -0.1874\\
%\multicolumn{1}{l}{}\\
& 2p$_{1/2}$ &                             &-0.3613\\
&	         &				               &	     \\
& 1f$_{5/2}$ & \multirow{2}{*}{ $^{49}$Ti$_{3.469}$ (1/2$^-$)} &   0.4981\\
& 2p$_{3/2}$ &                             & -0.3551\\
&	         &				               &	     \\
& 1f$_{7/2}$ & \multirow{4}{*}{ $^{49}$Ti$_{3.511}$ (5/2$^-$)} &   0.0081\\
& 1f$_{5/2}$ &                             & -0.0871\\
& 2p$_{3/2}$ &                             & -0.1451\\
& 2p$_{1/2}$ &                             & -0.0081\\
&	         &				               &	     \\
& 1f$_{7/2}$ & \multirow{1}{*}{ $^{49}$Ti$_{3.618}$ (5/2$^-$)} &  -0.0328\\\cline{1-4}
\multirow{13}{*}{$^{48}$Ti$_{0.984}$ (2$^+$)} & 1f$_{5/2}$ & \multirow{3}{*}{ $^{49}$Ti$_{3.618}$ (5/2$^-$)} &  0.0080\\
& 2p$_{3/2}$ &                             &  0.0608\\
& 2p$_{1/2}$ &                             & -0.1720\\
&	         &				               &	     \\
& 1f$_{7/2}$ & \multirow{4}{*}{ $^{49}$Ti$_{3.788}$ (3/2$^-$)} &  -0.0606\\
& 1f$_{5/2}$ &                             &  0.1221\\
& 2p$_{3/2}$ &                             & -0.1719\\
& 2p$_{1/2}$ &                             & -0.1221\\
&	         &				               &	     \\
& 1f$_{7/2}$ & \multirow{4}{*}{ $^{49}$Ti$_{3.855}$ (5/2$^-$)} &  -0.0671\\
& 1f$_{5/2}$ &                             & -0.0129\\
& 2p$_{3/2}$ &                             & -0.0764\\
& 2p$_{1/2}$ &                             & -0.2506\\\cline{1-4}
\multirow{26}{*}{$^{48}$Ti$_{3.359}$ (3$^-$)} & 1d$_{5/2}$ & \multirow{3}{*}{ $^{49}$Ti$_{g.s.}$ (7/2$^-$)} &  -0.0539\\
& 1d$_{3/2}$ &                             &  0.1277\\
& 2s$_{1/2}$ &                             & -0.1464\\
&	         &				               &	     \\
& 1d$_{5/2}$ & \multirow{2}{*}{ $^{49}$Ti$_{1.382}$ (3/2$^-$)} &   0.0160\\
& 1d$_{3/2}$ &                             & -0.0406\\
&	         &				               &	     \\
& 1d$_{5/2}$ & \multirow{2}{*}{ $^{49}$Ti$_{1.586}$ (3/2$^-$)} &  -0.0529\\
& 1d$_{3/2}$ &                             &  0.2503\\
&	         &				               &	     \\
& 1d$_{5/2}$ & \multirow{3}{*}{ $^{49}$Ti$_{1.623}$ (5/2$^-$)} &   0.0176\\
& 1d$_{3/2}$ &                             &  0.0151\\
& 2s$_{1/2}$ &                             &  0.0063\\
&	         &				               &	     \\
& 1d$_{5/2}$ & \multirow{1}{*}{ $^{49}$Ti$_{1.723}$ (1/2$^-$)} &   0.0024\\
&	         &				               &	     \\
& 1d$_{5/2}$ & \multirow{3}{*}{ $^{49}$Ti$_{1.762}$ (5/2$^-$)} &   0.0130\\
& 1d$_{3/2}$ &                             &  0.0258\\
& 2s$_{1/2}$ &                             &  0.0232\\
&	         &				               &	     \\
& 1d$_{5/2}$ & \multirow{3}{*}{ $^{49}$Ti$_{2.261}$ (5/2$^-$)} &  -0.0226\\
& 1d$_{3/2}$ &                             & -0.0596\\
& 2s$_{1/2}$ &                             & -0.0366\\
&	         &				               &	     \\
& 1d$_{5/2}$ & \multirow{2}{*}{ $^{49}$Ti$_{2.471}$ (5/2$^-$)} &  -0.0022\\
& 2s$_{1/2}$ &                             & -0.0032\\\cline{1-4}
\end{longtable}

%**************Table 1***************************
\setcounter{table}{3}
\begin{table}
\begin{center}
\caption{continued}
\begin{tabular}{|c| c| c| c|}
\hline
Initial state & \parbox{1.1cm}{\centering{\textbf{nlj}}} & \parbox{2.2cm}{\centering{Final state}} & \parbox{2.0cm}{\centering{Spectroscopic amplitude}}\\
\hline
\multirow{36}{*}{$^{48}$Ti$_{3.359}$ (3$^-$)} & 1d$_{5/2}$ & \multirow{3}{*}{ $^{49}$Ti$_{2.513}$ (5/2$^-$)} &   0.0125\\
& 1d$_{3/2}$ &                             & -0.0937\\
& 2s$_{1/2}$ &                             &  0.0353\\
&	         &				               &	     \\
& 1d$_{5/2}$ & \multirow{3}{*}{ $^{49}$Ti$_{2.980}$ (7/2$^-$)} &   0.0070\\
& 1d$_{3/2}$ &                             & -0.0203\\
& 2s$_{1/2}$ &                             &  0.0353\\
&	         &				               &	     \\
& 1d$_{5/2}$ & \multirow{3}{*}{ $^{49}$Ti$_{3.042}$ (7/2$^-$)} &  -0.0218\\
& 1d$_{3/2}$ &                             &  0.0458\\
& 2s$_{1/2}$ &                             & -0.0358\\
&	         &				               &	     \\
& 1d$_{5/2}$ & \multirow{1}{*}{ $^{49}$Ti$_{3.175}$ (1/2$^-$)} &   0.0341\\
&	         &				               &	     \\
& 1d$_{5/2}$ & \multirow{2}{*}{ $^{49}$Ti$_{3.261}$ (3/2$^-$)} &   0.0049\\
& 1d$_{3/2}$ &                             & -0.0253\\
&	         &				               &	     \\
& 1d$_{5/2}$ & \multirow{2}{*}{ $^{49}$Ti$_{3.428}$ (3/2$^-$)} &  -0.0085\\
& 1d$_{3/2}$ &                             &  0.0908\\
&	         &				               &	     \\
& 1d$_{5/2}$ & \multirow{1}{*}{ $^{49}$Ti$_{3.469}$ (1/2$^-$)} &  -0.0059\\
&	         &				               &	     \\
& 1d$_{5/2}$ & \multirow{3}{*}{ $^{49}$Ti$_{3.511}$ (5/2$^-$)} &   0.0052\\
& 1d$_{3/2}$ &                             & -0.0290\\
& 2s$_{1/2}$ &                             & -0.0032\\
&	         &				               &	     \\
& 1d$_{5/2}$ & \multirow{3}{*}{ $^{49}$Ti$_{3.618}$ (5/2$^-$)} &  -0.0062\\
& 1d$_{3/2}$ &                             &  0.0192\\
& 2s$_{1/2}$ &                             & -0.0038\\
&	         &				               &	     \\
& 1d$_{5/2}$ & \multirow{2}{*}{ $^{49}$Ti$_{3.788}$ (3/2$^-$)} &  -0.0120\\
& 1d$_{3/2}$ &                             &  0.0746\\
&	         &				               &	     \\
& 1d$_{5/2}$ & \multirow{3}{*}{ $^{49}$Ti$_{3.855}$ (5/2$^-$)} &   0.0076\\
& 1d$_{3/2}$ &                             &  0.0013\\
& 2s$_{1/2}$ &                             &  0.0026\\
\hline
\end{tabular}
\end{center}
\end{table}

%
%
%
%**************Table 5***************************
\begin{table*}
\renewcommand\thetable{V}
\begin{center}
\caption{One-neutron transfer cross-sections integrated in the angular range [5$^{o}$-14$^{o}$] in the centre of mass reference frame for each ROI in Fig. 6b. The populated states which contribute to the overall cross-section at each ROI are reported in the second column. The theoretical cross-sections for the individual transitions were calculated under the DWBA and CCBA frameworks using the psdmod and KB3 interactions and are presented in the third, fourth and fifth column. The sum of the predicted cross-sections for all transitions contributing to each ROI, $\sigma_{tot.}$, is given in the sixth, seventh and eighth column and is compared to the experimental value, $\sigma_{exp.}$, presented in the last column.}
\begin{tabular}{| c| c |c |c |c |c |c |c |c |}
\hline
\hline
ROI & \parbox{2.5cm}{\centering{Populated states}} & \parbox{2.0cm}{\centering{$\sigma_{DWBA}^{psdmod}$ (mb)}} & \parbox{2.0cm}{\centering{$\sigma_{CCBA}^{psdmod}$ (mb)}} & \parbox{2.0cm}{\centering{$\sigma_{DWBA}^{KB3}$ (mb)}} & \parbox{2.2cm}{\centering{$\sigma_{tot.}^{DWBA-psdmod}$ (mb)}} & \parbox{2.2cm}{\centering{$\sigma_{tot.}^{CCBA-psdmod}$ (mb)}} & \parbox{2.2cm}{\centering{$\sigma_{tot.}^{DWBA-KB3}$ (mb)}} & \parbox{1.8cm}{\centering{$\sigma_{exp.}$ (mb)}}\\
\hline
1 & $^{17}$O$_{g.s.}$+$^{49}$Ti$_{g.s.}$ & 0.529  & 0.637 & 0.547 &- &- &- &0.446 $\pm$ 0.015\\\cline{1-9}
\multirow{7}{*}{2} & $^{17}$O$_{g.s.}$+$^{49}$Ti$_{1.382}$ & 1.448  & 1.409 & 1.218 &\multirow{7}{*}{1.890} &\multirow{7}{*}{1.865} &\multirow{7}{*}{1.579} &\multirow{7}{*}{1.057 $\pm$ 0.031}\\
                   & $^{17}$O$_{g.s.}$+$^{49}$Ti$_{1.586}$ & 0.028  & 0.027 & 0.057 & & & &\\
                   & $^{17}$O$_{g.s.}$+$^{49}$Ti$_{1.623}$ & 0.009  & 0.014 & 0.003 & & & &\\
                   & $^{17}$O$_{g.s.}$+$^{49}$Ti$_{1.723}$ & 0.135  & 0.146 & 0.144 & & & &\\
                   & $^{17}$O$_{g.s.}$+$^{49}$Ti$_{1.762}$ & 0.183  & 0.173 & 0.087 & & & &\\
                   & $^{17}$O$_{g.s.}$+$^{49}$Ti$_{2.261}$ & 0.024  & 0.022 & 0.005 & & & &\\
                   & $^{17}$O$_{0.871}$+$^{49}$Ti$_{g.s.}$ & 0.063  & 0.074 & 0.065 & & & &\\\cline{1-9}
\multirow{15}{*}{3} & $^{17}$O$_{g.s.}$+$^{49}$Ti$_{1.723}$ & 0.015  & 0.016 & 0.016 & \multirow{15}{*}{0.801} & \multirow{15}{*}{0.865} & \multirow{15}{*}{0.499} & \multirow{15}{*}{0.563 $\pm$ 0.020} \\
                   & $^{17}$O$_{g.s.}$+$^{49}$Ti$_{1.762}$ & 0.029  & 0.027 & 0.014 & & & &\\
                   & $^{17}$O$_{g.s.}$+$^{49}$Ti$_{2.261}$ & 0.184  & 0.166 & 0.005 & & & &\\
                   & $^{17}$O$_{g.s.}$+$^{49}$Ti$_{2.471}$ & 0.047  & 0.040 & 0.026 & & & &\\
                   & $^{17}$O$_{g.s.}$+$^{49}$Ti$_{2.513}$ & 0.049  & 0.069 & 0.009 & & & &\\
                   & $^{17}$O$_{g.s.}$+$^{49}$Ti$_{2.980}$ & 0.040  & 0.064 & 0.020 & & & &\\
                   & $^{17}$O$_{g.s.}$+$^{49}$Ti$_{3.042}$ & 0.022  & 0.028 & 0.028 & & & &\\
                   & $^{17}$O$_{g.s.}$+$^{49}$Ti$_{3.175}$ & 0.016  & 0.018 & 0.022 & & & &\\
                   & $^{17}$O$_{g.s.}$+$^{49}$Ti$_{3.261}$ & 0.012  & 0.014 & 0.029 & & & &\\
                   & $^{17}$O$_{0.871}$+$^{49}$Ti$_{1.382}$ & 0.243  & 0.260 & 0.204 & & & &\\
                   & $^{17}$O$_{0.871}$+$^{49}$Ti$_{1.586}$ & 0.006  & 0.007 & 0.011 & & & &\\
                   & $^{17}$O$_{0.871}$+$^{49}$Ti$_{1.623}$ & 0.002  & 0.003 & $<$0.001 & & & &\\
                   & $^{17}$O$_{0.871}$+$^{49}$Ti$_{1.723}$ & 0.048  & 0.053 & 0.051 & & & &\\
                   & $^{17}$O$_{0.871}$+$^{49}$Ti$_{1.762}$ & 0.048  & 0.050 & 0.023 & & & &\\
                   & $^{17}$O$_{3.055}$+$^{49}$Ti$_{g.s.}$  & 0.040  & 0.050 & 0.041 & & & &\\\cline{1-9}
\multirow{21}{*}{4} & $^{17}$O$_{g.s.}$+$^{49}$Ti$_{2.980}$ & 0.035  & 0.055 & 0.017 &\multirow{21}{*}{0.562} &\multirow{21}{*}{0.633} &\multirow{21}{*}{0.655} &\multirow{21}{*}{0.645 $\pm$ 0.027}\\
                   & $^{17}$O$_{g.s.}$+$^{49}$Ti$_{3.042}$ & 0.031  & 0.038 & 0.039 & & & &\\
                   & $^{17}$O$_{g.s.}$+$^{49}$Ti$_{3.175}$ & 0.060  & 0.068 & 0.084 & & & &\\
                   & $^{17}$O$_{g.s.}$+$^{49}$Ti$_{3.261}$ & 0.095  & 0.108 & 0.225 & & & &\\
                   & $^{17}$O$_{g.s.}$+$^{49}$Ti$_{3.428}$ & 0.004  & 0.007 & 0.021 & & & &\\
                   & $^{17}$O$_{g.s.}$+$^{49}$Ti$_{3.469}$ & 0.087  & 0.064 & 0.071 & & & &\\
                   & $^{17}$O$_{g.s.}$+$^{49}$Ti$_{3.511}$ & 0.012  & 0.012 & $<$0.001 & & & &\\
                   & $^{17}$O$_{g.s.}$+$^{49}$Ti$_{3.618}$ & 0.010  & 0.013 & 0.025 & & & &\\
                   & $^{17}$O$_{g.s.}$+$^{49}$Ti$_{3.788}$ & 0.002  & 0.004 & $<$0.001 & & & &\\
                   & $^{17}$O$_{g.s.}$+$^{49}$Ti$_{3.855}$ & 0.014  & 0.018 & 0.005 & & & &\\
                   & $^{17}$O$_{0.871}$+$^{49}$Ti$_{2.261}$ & 0.038  & 0.036 & $<$0.001 & & & &\\
                   & $^{17}$O$_{0.871}$+$^{49}$Ti$_{2.471}$ & 0.012  & 0.010 & 0.006 & & & &\\
                   & $^{17}$O$_{0.871}$+$^{49}$Ti$_{2.513}$ & 0.012  & 0.017 & 0.002 & & & &\\
                   & $^{17}$O$_{0.871}$+$^{49}$Ti$_{2.980}$ & 0.012  & 0.017 & 0.006 & & & &\\
                   & $^{17}$O$_{0.871}$+$^{49}$Ti$_{3.042}$ & 0.008  & 0.009 & 0.010 & & & &\\
                   & $^{17}$O$_{0.871}$+$^{49}$Ti$_{3.175}$ & 0.012  & 0.014 & 0.017 & & & &\\
                   & $^{17}$O$_{0.871}$+$^{49}$Ti$_{3.261}$ & 0.007  & 0.009 & 0.017 & & & &\\
                   & $^{17}$O$_{0.871}$+$^{49}$Ti$_{3.428}$ & $<$0.001  & $<$0.001 & $<$0.001 & & & &\\
                   & $^{17}$O$_{0.871}$+$^{49}$Ti$_{3.469}$ & 0.003  & 0.002 & 0.002 & & & &\\
                   & $^{17}$O$_{3.055}$+$^{49}$Ti$_{g.s.}$ & 0.073  & 0.092 & 0.075 & & & &\\
                   & $^{17}$O$_{3.055}$+$^{49}$Ti$_{1.382}$ & 0.035  & 0.040 & 0.030 & & & &\\
                   \hline
\end{tabular}
\end{center}
\end{table*}
\end{document}